\documentclass[journal=aaemcq,manuscript=article]{achemso}
\usepackage{orcidlink}
\usepackage{bm}
\usepackage{mathptmx}
\usepackage{epstopdf}
\usepackage{booktabs}
\usepackage{multirow}
\usepackage{hhline}
\usepackage{dcolumn}
\usepackage{subfigure}
\usepackage{amsmath}
\usepackage{float}
\usepackage{chemformula}
\usepackage[version=3]{mhchem}
\usepackage{chemfig}
\usepackage[autostyle]{csquotes}
\usepackage{soul}
\usepackage{ulem} 
\usepackage{xr}
\usepackage{cleveref}

\title{A Hybrid Machine Learning Framework for Predicting Hydrogen Storage Capacities in Metal Hydrides: Unsupervised Feature Learning with Deep Neural Networks}
\author{Satadeep Bhattacharjee}
\email{s.bhattacharjee@ikst.res.in}
\affiliation{Indo-Korea Science and Technology Center (IKST), Jakkur, Bengaluru 560065, India}
\author{Pritam Das}
\author{Swetarekha Ram}
\affiliation{Indo-Korea Science and Technology Center (IKST), Jakkur, Bengaluru 560065, India}
\author{Seung-Cheol Lee}
\email{leesc@kist.re.kr}
\affiliation{Electronic Materials Research Center, Korea Institute of Science and Technology, Seoul 02792, Republic of Korea}


\begin{document}

\begin{abstract}
In this study, we present a sophisticated hybrid machine-learning framework that significantly improves the accuracy of predicting hydrogen storage capacities in metal hydrides. This is a critical challenge due to the scarcity of experimental data and the complexity of high-dimensional feature spaces. Our approach employs the power of unsupervised learning through the use of a state-of-the-art autoencoder. This autoencoder is trained on elemental descriptors obtained from Mendeleev software, enabling the extraction of a meaningful and lower dimensional latent space from the input data. This latent representation serves as the basis for our deep multi-layer perceptron (MLP) model, which consists of five layers and shows good precision in predicting hydrogen storage capacities. Furthermore, our results show very good agreement with the results obtained with density functional theory (DFT). In addition to addressing the limitations caused by limited and unevenly distributed data in the field of hydrogen storage materials, we also focus on discovering new materials that show promising opportunities for hydrogen storage. These materials were identified using both feature-based approaches and predictions generated by a large language model (LLM). A significant highlight of this work is the discovery of new hydrogen storage materials using a LLM, with a selected subset subsequently validated through density functional theory (DFT) calculations. Finally, our investigation into the effectiveness of transferring weights from the autoencoder to the MLP, in addition to the latent features, suggests that while this strategy slightly improves model performance as indicated by a slightly higher R$^2$ value and lower RMSE, it emphasizes the intricate challenge of adapting pre-trained weights for specific supervised tasks.
\end{abstract}
\maketitle
\section{Introduction}
Climate change, energy security and the potential depletion of resources due to growing global population and technological advances represent pressing challenges for humanity.~\cite{zuttel2003materials, Wonder}. To address these issues, promoting renewable energy is crucial as hydrogen is becoming an essential energy source for mobile and stationary applications. Hydrogen not only reduces environmental damage, but also reduces dependence on imported oil for countries without natural resources~\cite{barthelemy2017hydrogen}. There are various storage technologies for hydrogen, including compressed gas, cryogenic liquids and solid fuels as a chemical or physical combination with other materials such as metal hydrides, complex hydrides and carbon materials or produced in vehicles by on-site methanol reforming~\cite{oesterreicher1981hydrides,lee2005review}. Each of these methods possesses unique attributes for hydrogen storage. Storage by absorption as chemical compounds or by adsorption on carbon materials carries safety benefits, necessitating a form of conversion or energy input for hydrogen release. Considerable effort has been invested into new hydrogen-storage systems, including metal, chemical, or complex hydrides, and carbon nanostructures~\cite{Norskov,lai2015hydrogen}.

Hydrogen storage in metal hydrides involves chemically combining hydrogen gas with metal elements, forming metal hydrides. These offer significant safety benefits over gas and liquid storage methods, while also boasting higher hydrogen storage density (6.5 H atoms/cm\textsuperscript{3} for MgH\textsubscript{2}) as compared to hydrogen gas (0.99 H atoms/cm\textsuperscript{3}) or liquid hydrogen (4.2 H atoms/cm\textsuperscript{3}) \cite{astle1974crc}. Hence, metal hydride storage is a safe, volume-efficient method suitable for onboard vehicle applications. Hydrogen typically stored within the lattice structure of metal hydrides is attached to metal atoms. Metal-hydrogen combinations yield two types of hydrides, $\alpha$-phase which absorbs some hydrogen, and $\beta$-phase where hydride is fully formed \cite{GRAY20111630}. Hydrogen storage in metal hydrides is influenced by various factors and involves several mechanistic steps. The ability of metals to dissociate hydrogen depends on surface structure, morphology, and purity \cite{DAVID2005169}. Factors such as the type of metal, type of hydride, and operating conditions impact the hydrogen storage capacity of metal hydrides. Light metals like Li, Be, Na, Mg, B, and Al form numerous metal–hydrogen compounds and are particularly intriguing owing to their lightweight and the large number of hydrogen atoms per metal atom, often in the order of H/M = 2 \cite{VERMA2023, das2023computational, LI201114512, ALI20211111, BARTHELEMY20177254}.

Many metals have the capacity to engage with hydrogen in order to produce binary hydrides (MH\textsubscript{n}). Nonetheless, the majority of binary hydrides lack the necessary properties for effective hydrogen storage as a carrier. Certain intermetallic compounds have the ability to generate hydrides with structural formulas of AB\textsubscript{x}H\textsubscript{n}. In this scenario, element A, which typically belongs to the rare earth or alkaline earth metal category due to its strong hydrogen affinity, forms a stable hydride, while element B tends to create only unstable hydrides given its low hydrogen affinity \cite{WESTLAKE19831, oesterreicher1981hydrides}. The metal hydrides that have been extensively examined are those falling under AB\textsubscript{2} and AB\textsubscript{5}, particularly in relation to hydrogen storage and applications in fuel cells \cite{LOTOTSKYY20173}. Intermetallic compounds of the AB\textsubscript{5} type stand out due to their easily achievable activation, swift kinetics in hydrogen absorption and release, as well as the relatively high stability exhibited in hydrogen sorption properties throughout cyclic hydrogenation/dehydrogenation processes \cite{LOTOTSKYY20145818}. Pressure-composition isotherms in H-AB\textsubscript{5} systems show a single flat plateau with low H\textsubscript{2} absorption-desorption hysteresis, which increases in substituted alloys \cite{TARASOV20184415}. Ti and Zr are the most common A components in AB\textsubscript{2}-type compounds, and the B component is usually represented by a transition metal such as Mn, Cr, Fe, or V \cite{LOTOTSKYY20145818}. AB\textsubscript{2}-type alloys are generally less easily activated than AB\textsubscript{5} alloys, and they can be doped with small amounts ($\sim$1 at\%) of rare earth elements to facilitate activation \cite{YAO2018524}. They have excellent hydrogenation/dehydrogenation kinetics and cycle stability once activated \cite{SANDROCK1999877}. The AB\textsubscript{2} components are less expensive than the AB\textsubscript{5} components, but manufacturing of the AB\textsubscript{2}-type alloys presents some metallurgical challenges due to higher melting temperatures, high component reactivity, and other factors \cite{SANDROCK1999877, FASHU2020108295}. The most studied metal hydrides for hydrogen storage include sodium aluminum hydride, magnesium hydride, aluminum hydride, LaNi\textsubscript{5}, and ZrV\textsubscript{2} \cite{NIAZ2015457, WEI20171122, TARASOV20184415, ZOTOV2008220}.

The introduction of data-driven materials design~\cite{dd1,dd2,dd3} has accelerated material discovery, processing, and manufacturing \cite{SPARKS201610, RAHNAMA2018169, ramprasad2017machine,louis2022accurate,bhattacharjee2022general}. Machine learning (ML) algorithms allow for the creation of new alloys based solely on previously collected data, either publicly available or reported in scientific literature. Metal hydrides are an ideal subject for machine learning algorithm research due to their diverse properties and broad range. Traditional methods for determining the optimized materials class and corresponding metal hydride composition based on desirable properties are difficult, time-consuming, and costly. ML techniques, on the other hand, enable rapid, productive, and efficient material class prediction for a specific hydrogen weight percent and operational conditions. Numerous groups have begun to use ML techniques to speed up this screening and gain more physical insight from massive amounts of data \cite{ahmed2021predicting, ALI2022105844, RAHNAMA20197337}. Rezakazemi et al. used an adaptive neuro-fuzzy inference system to evaluate the performance of hydrogen-selective mixed matrix membranes under various operational conditions \cite{REZAKAZEMI201715211}. Rahnama et al. predicted the hydrogen storage capacities in various metal hydrides using machine learning algorithms in another study and found that higher temperatures yielded higher hydrogen storage capacities \cite{RAHNAMA20197337, RAHNAMA20197345}. Ahmed et al. recently predicted gravimetric and volumetric hydrogen capacities in metal-organic frameworks (MOFs) using ML algorithms \cite{ahmed2019exceptional}. ML methods have been instrumental in predicting various properties of (MOFs) and even designing new structures. This approach has significantly changed the research landscape in this field, allowing for the rapid identification of promising MOF candidates~\cite{MOF2,MOF3} for various applications, including hydrogen storage.
In a recent work, Sun \textit{et al.}  introduced a method that combines meta-learning and high-throughput molecular simulations to predict hydrogen storage in nanoporous materials efficiently.~\cite{sun2019predicting}

In this study, we propose a two-stage model for predicting hydrogen storage capacity in metal hydrides. The first stage involves unsupervised learning to extract latent features from the input data, and the second stage employs a multi-layer perceptron (MLP) trained to predict target properties. We developed a deep neural network-based model that can predict the hydrogen storage capacities in metal hydrides. We also identified new hydrogen storage materials using both feature-based approaches as well as using a large language model (LLM). The hydrogen storage capacity of these predicted materials was calculated using the above-mentioned approach and further compared with the results from the DFT-based methods. A chemical bonding analysis was performed to comprehend the relationship between storage capacity and the atomic environment, providing insights into the chemical bonds in the materials and their contribution to hydrogen storage capacity.

It is worth noting that the application of LLMs in materials science has shown significant potential as they leverage their advanced natural language processing capabilities to predict and design novel materials. Recent advances include the development of the Materials Informatics Transformer (MatInFormer), which leverages tokenization of crystallographic space group information for high-precision predictions of material properties, particularly in metal-organic frameworks (MOFs).~\cite{Huang2023Materials}. Furthermore, transformer models such as GPT and BART have been successfully applied to generative design tasks, producing chemically valid and novel material compositions with a high degree of accuracy in terms of charge neutrality and electronegativity balance~\cite{fu2023material}. As another example, the HoneyBee model illustrates the specific adaptation of LLMs for materials science through iterative fine-tuning processes~\cite{Song2023HoneyBee:} and demonstrates superior performance on domain-specific tasks.
\section{Methods}
\begin{figure}[htbp]
	\centering
	\includegraphics[width=1.0\linewidth]{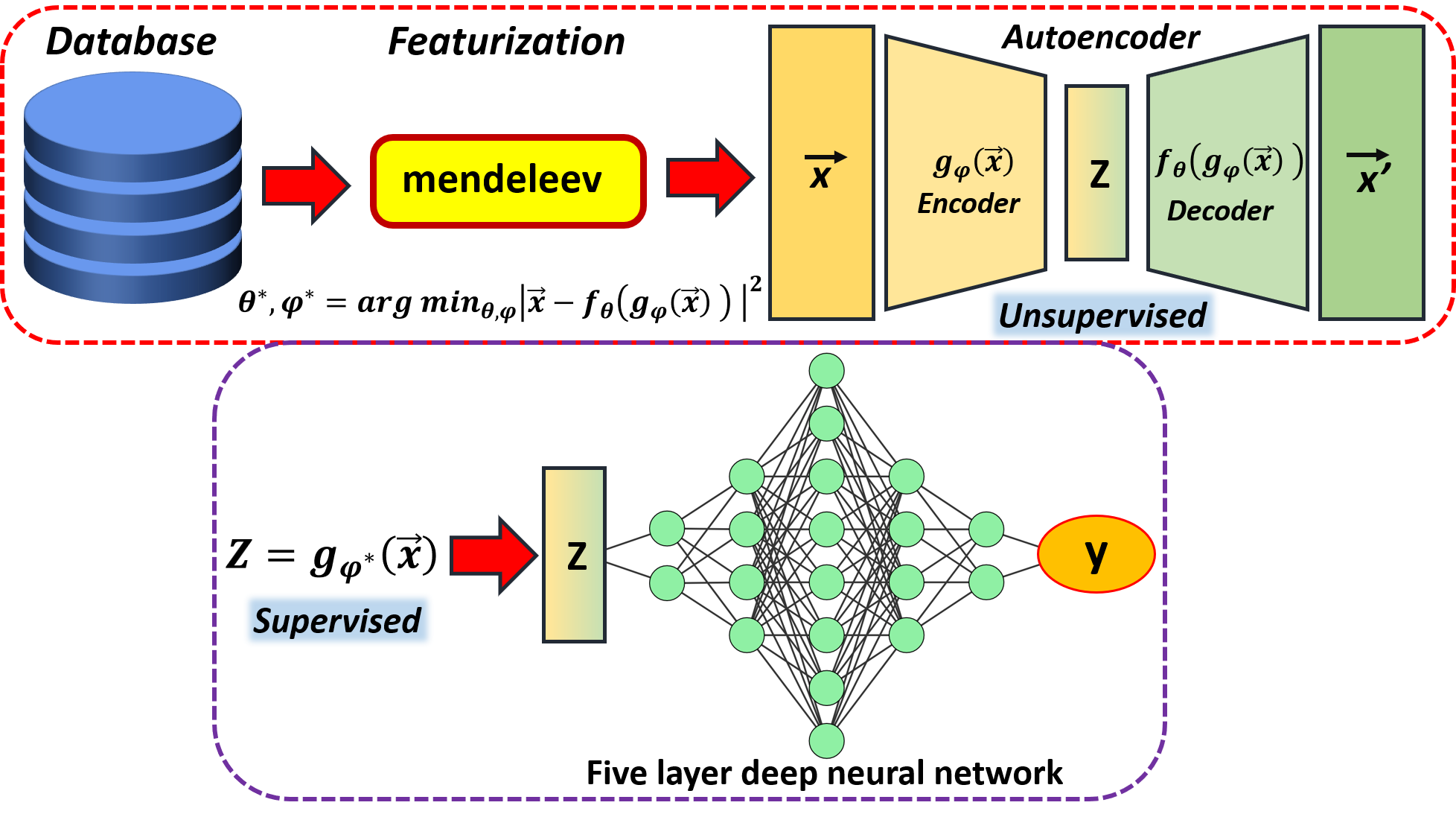}
	\caption{Architecture  of our hybrid approach}
	\label{Schematic}
\end{figure}

The architecture of our method is shown in Fig. \ref{Schematic}, in a schematic way. We use a hybrid approach that has two parts: in the unsupervised learning part, an autoencoder is trained on the entire dataset, which consists of only unlabeled data, to learn a compact representation of the data in the bottleneck layer \cite{8616075}. A Multi-Layer Perceptron (MLP) is used in the supervised learning part to predict the hydrogen storage capacity. The major use of such an approach is to address the challenges of data scarcity and high feature dimensionality. In this study, the initial feature space has a high dimensionality and the available dataset is relatively small. As there are not too many machine learning studies available in the literature, it is not easy to choose proper descriptors. We have constructed 36 features using the Mendeleev software \cite{mendeleev2014}. To handle high-dimensional feature spaces with small datasets, we used an autoencoder to reduce the dimensionality of the data. 

\subsection{Unsupervised learning}
In the unsupervised learning part, we train an autoencoder on the entire dataset, which consists of only unlabeled data. The goal of the autoencoder is to learn a compact representation (or feature representation) of the data in the bottleneck layer. 

Autoencoders are a kind of neural network that is used to extract features in an unsupervised manner. They are divided into two parts: an encoder, which compresses the input data into a lower-dimensional representation (also known as the \textit{latent representation} or \textit{bottleneck}), and a decoder, which reconstructs the original input from the compressed form \cite{lecun2015deep}.

The autoencoder's purpose during training is to minimize the reconstruction error between the input and its reconstruction. This forces the encoder to learn a compact representation of the input data that maintains just the most significant information, which can then be utilized to extract features. Once trained, the encoder may be used to extract features from new, previously unseen data by feeding it through the network and using the bottleneck layer activations as features.

Let $\vec{x}$ be the input vector, and let $g_{\phi}$ and $f_{\theta}$ represent the encoder and decoder functions, respectively. The encoder function maps the input vector to a lower-dimensional representation $z$:

\begin{equation}
z = g_{\phi}(\vec x) = \sigma(W_{\phi}\vec x + b_{\phi})
\end{equation}

where $\sigma$ is the activation function, $W_{\phi}$ and $b_{\phi}$ are the weight and bias parameters of the encoder, respectively. For the activation function, we have used the RelU function as implemented in PyTorch \cite{NEURIPS2019_9015}. 

The decoder function maps the lower-dimensional representation back to the original input space:

\begin{equation}
\hat{x} = f_{\theta}(z) = \sigma(W_{\theta}z + b_{\theta})
\end{equation}

where $W_{\theta}$ and $b_{\theta}$ are the weight and bias parameters of the decoder, respectively.

The goal of the autoencoder is to minimize the reconstruction error, defined as the difference between the original input $x$ and its reconstruction $\hat{x}$,

\begin{equation}
\mathcal{L}(\phi, \theta) = ||\vec x - \hat{x}||^2
\end{equation}

This loss function was minimized using the Adam optimization method to find the optimal values for the parameters $\phi$ and $\theta$ say $\phi^*$ and $\theta^*$  \cite{kingma2014adam}. Once the autoencoder is trained, the encoder function $z=g_{\phi^*}(x)$ can be used to extract features from the input data by using the activations of the lower-dimensional representation $z$ as the features.

The network architecture for both the encoder and decoder includes two linear layers and dropout layers to prevent overfitting. The input to the autoencoder is first passed through a linear layer with 256 units, followed by a Leaky ReLU activation function \cite{ledig2017photo}. This is then followed by a dropout layer with a dropout probability of 0.3 to provide regularization. 

As shown in Fig. \ref{Schematic}, we create the elemental features using the Mendeleev software. These features are then fed into the autoencoder to learn latent information in an unsupervised manner. The importance of the latent space generation step lies in its ability to learn a more compact and relevant representation of the input features, potentially highlighting complex, nonlinear relationships between them that might be difficult to capture with a simple MLP. By training the MLP on these learned features instead of the original features helps to improve the prediction accuracy of the model. Also, as the dimension of the original feature is large it is very difficult to work with an MLP which is very which has five layers.

Given the unsupervised nature of the autoencoder training, it has the additional benefit of being less prone to overfitting, and more robust to noise and outliers in the input data, compared to directly training a supervised model on the raw input features.

\subsection{Supervised learning}
Let \( z_i \) be the input data, and let \( y_i \) be the corresponding target (true) value, and \( L \) is the number of labeled data points.

We train a regressor, such as a Multilayer Perceptron (MLP), on the labeled data using the extracted features through the aforementioned method as inputs. The MLP takes these latent features as input and predicts the target values based on the labeled data. The loss function to minimize can be expressed as:
\begin{equation}
\mathcal{L} = \frac{1}{L} \sum_{i=1}^{L} (y_i - \hat{y}_i)^2
\end{equation}

Here, \( \hat{y}_i \) represents the predicted values from the MLP, and \(\mathcal{L}\) is the mean squared error loss over the labeled data points.

\subsection{\textit{ab-initio} calculations}
First-principles calculations were conducted employing the density functional theory (DFT) within the Vienna Ab initio Software Package (VASP) \cite{Kresse1996CMS, Kresse1996PRB}. To determine the electronic energy, projected augmented wave (PAW) potentials and plane-wave basis sets were utilized for both core and valence electrons, employing the Perdew–Burke–Ernzerhof (PBE) functional \cite{Kresse1999, Bl1994}. The calculations accounted for long-range interactions using the semi-empirical Grimme D2 dispersion method and incorporated non-spherical contributions to the PAW potentials within the code \cite{Grimme2006}. A cutoff energy of 500 eV ensured convergence of all energies. Structural optimization employed the conjugate gradient algorithm \cite{Perdew1996}, with convergence criteria set at 10\textsuperscript{-7} eV for energy and 0.01 eV\AA\textsuperscript{-1} for force. Spin polarization was enabled in all calculations, except for the isolated H\textsubscript{2} closed shell molecule.

\section{Dataset}

\begin{figure}[htbp]
	\centering
	\includegraphics[width=0.65\linewidth]{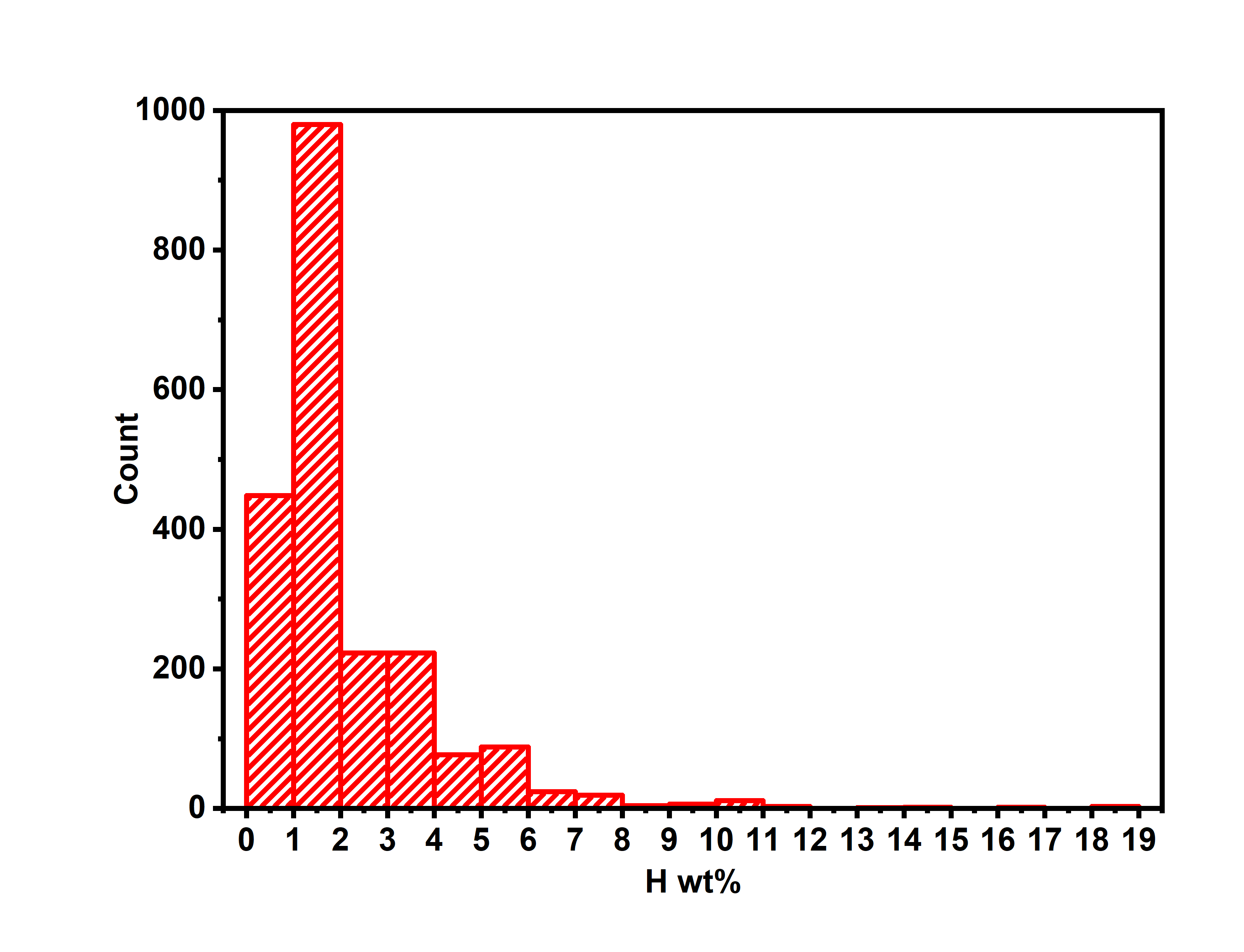}
	\caption{Histogram showing the hydrogen weight percentage distribution in the database used for the study.}
	\label{fig:data}
\end{figure}

For this study, we utilized the Hydrogen Storage Materials Database shared by the HyMARC Data Hub (https://datahub.hymarc.org/). This is an openly available database that can be accessed at https://datahub.hymarc.org/en/dataset/hydrogen-storage-materials-db. This database contains information on more than 2000 hydride materials and their properties. The Hydride Database classifies the hydrides into eight classes: A\textsubscript{2}B intermetallic compounds, AB intermetallic compounds, AB\textsubscript{2} intermetallic compounds, AB\textsubscript{5} intermetallic compounds, complex hydrides, Mg alloys, solid solution alloys, and Misc. In the dataset, more than 950 compounds have 1-2 wt\% hydrogen storage capacity (Fig. \ref{fig:data}). The average hydrogen storage capacity of all compounds is 2.1 wt\%.  The dataset has as high as 20.8 wt\% hydrogen storage capacity complex hydrides, which is BeB\textsubscript{4}H\textsubscript{8} \cite{doi:https://doi.org/10.1002/0471238961.0825041819211212.a01}.

\section{Description of the features}
\begin{figure}[htbp]
	\centering
	\includegraphics[width=1.0\linewidth]{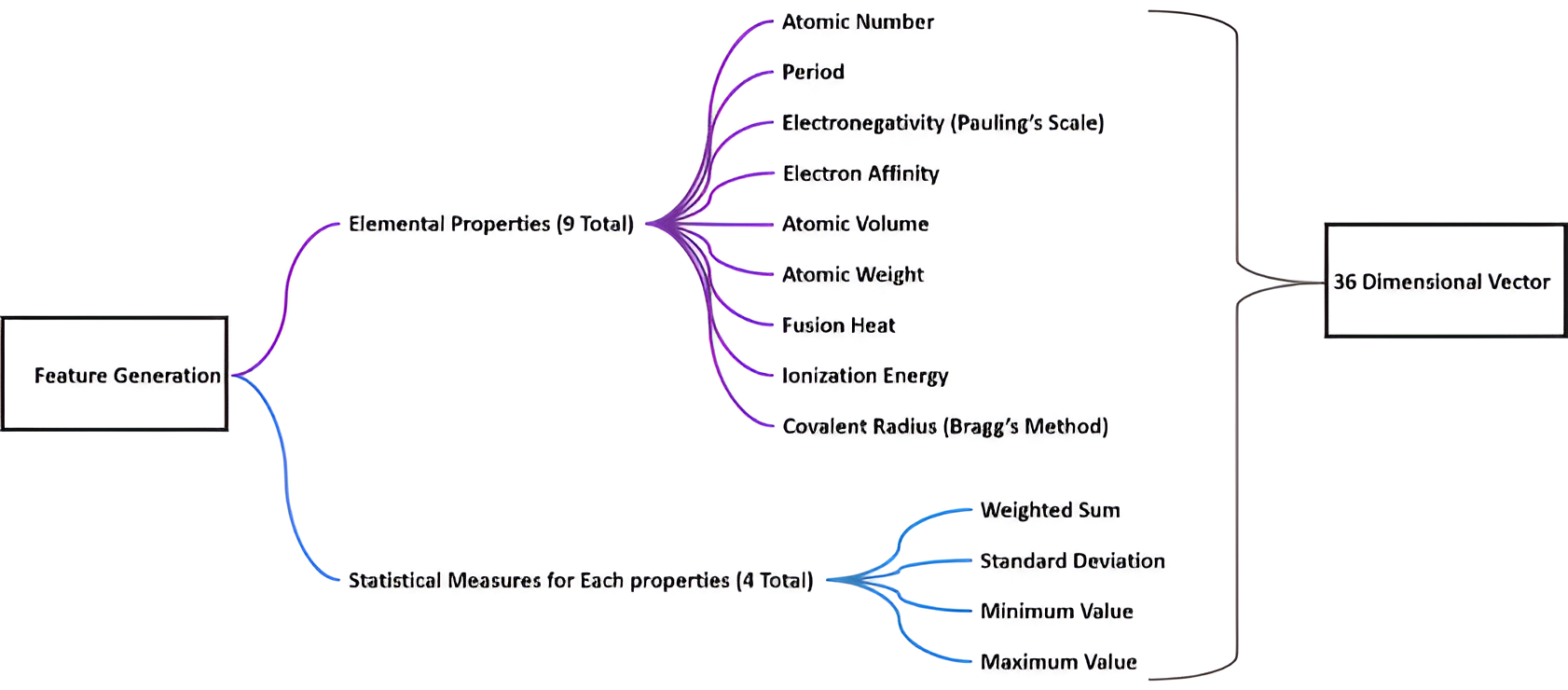}
	\caption{Architecture of feature generation algorithm.}
	\label{feature}
\end{figure}

For every chemical composition, our feature generation algorithm creates a 36-dimensional vector that contains a set of nine elemental properties (Fig. \ref{feature}). These characteristics include the atomic number, period, electronegativity (measured according to Pauling's scale), electron affinity, atomic volume, atomic weight, fusion heat, and ionization energy. The covalent radius is calculated using Bragg's method. Four statistical measures—the weighted sum, standard deviation, minimum, and maximum values—are computed within the compound's composition for each property, resulting in \( 9 \times 4 = 36 \) dimensions. 

\section{Results and Discussion}

\subsection{Machine Learning Model for the Prediction of Hydrogen Storage Capacity}

In this study, our objective was to devise an effective machine learning model for predicting hydrogen storage capacity. Our investigation was bifurcated into two distinct scenarios: firstly, employing a Multilayer Perceptron (MLP) trained directly on all the features, and secondly, utilizing a hybrid approach.

\subsection{Case I: Using the MLP Directly for All the Features}

\begin{figure}[htbp]
    \centering
    \includegraphics[width=0.85\linewidth]{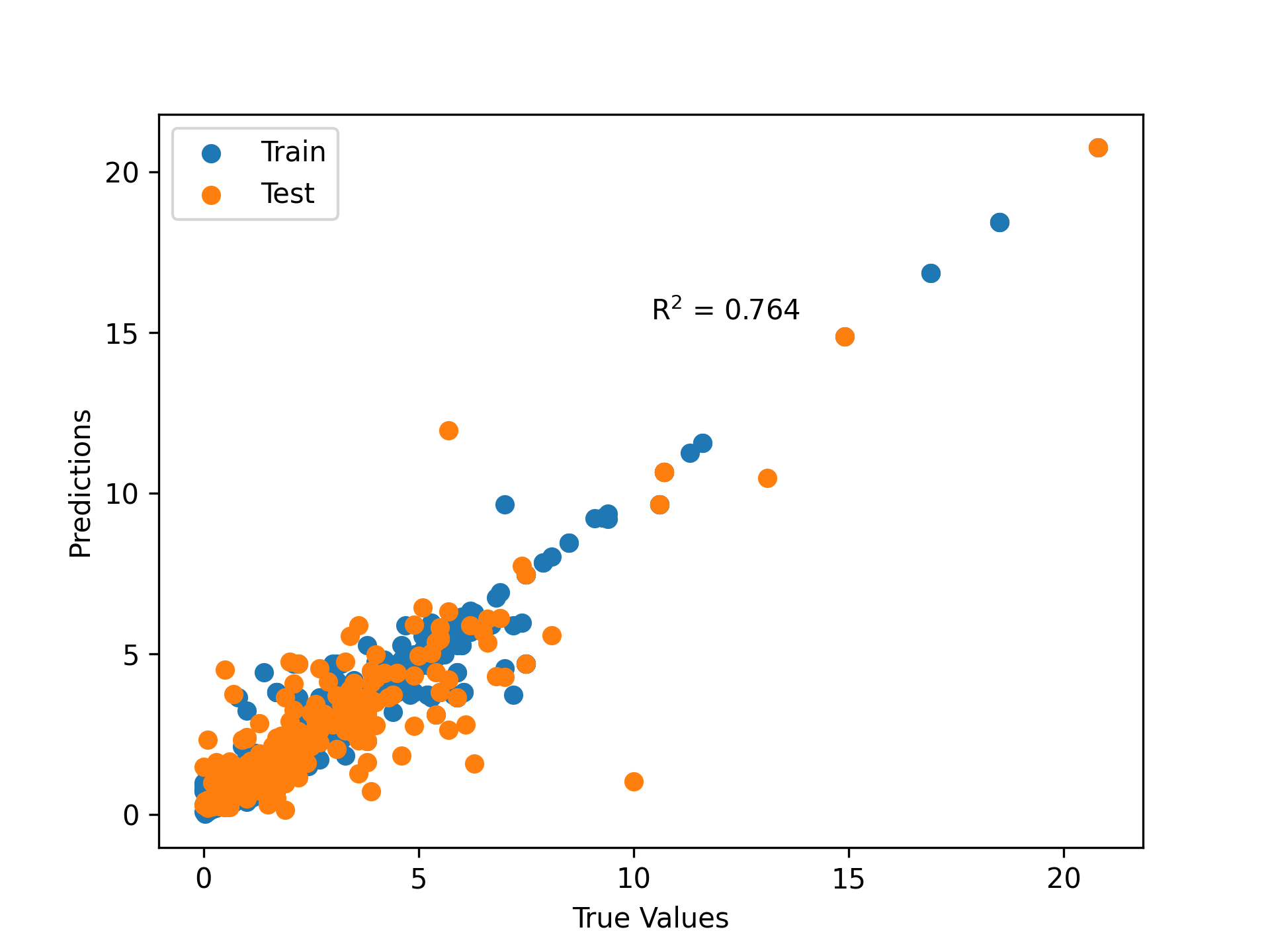}
    \caption{Pair plot contrasting the true values (from the dataset) with the predicted outcomes (both test and train) when the raw (36) features set is employed to train and test the MLP.}
    \label{fig:All-features}
\end{figure}

For our primary experiment, the MLP was trained and tested using all of the 36 features derived from the Mendeleev software. The dataset was strategically partitioned into training and testing subsets, with respective ratios of 0.8 and 0.2. The resultant outcomes of this methodology are visually represented in Fig.\ref{fig:All-features}. The predictive efficacy, quantified by the \(R^2\) coefficient, was deemed unsatisfactory. The coefficient of determination (\( R^2 \)) is 0.764, though a reasonably good fit, indicating that only approximately 76.4\% of the variance in the true values can be explained by the model.
Such results emphasize the inherent difficulties connected with managing datasets with a high number of dimensions, especially when the amount of data is small.

\subsection{Case II: Adopting the hybrid Approach}
Given the large number of dimensions present in the initial feature space (36 features) in contrast to the constrained size of the dataset, we opted to employ a hybrid approach. Our primary approach involved implementing an autoencoder to produce a latent space, which subsequently served as the input for a five-layer neural network (MLP). The objective of this downstream MLP was to predict the hydrogen storage capacity.

The optimal dimensions for the latent space were determined by assessing the model's performance across various latent space sizes. The dataset was partitioned into training and testing datasets, adopting a 0.8 and 0.2 split ratio. The training process utilized a learning rate of 0.001 and spanned 1000 epochs.

\begin{figure}[htbp]
    \centering
    \includegraphics[width=0.85\linewidth]{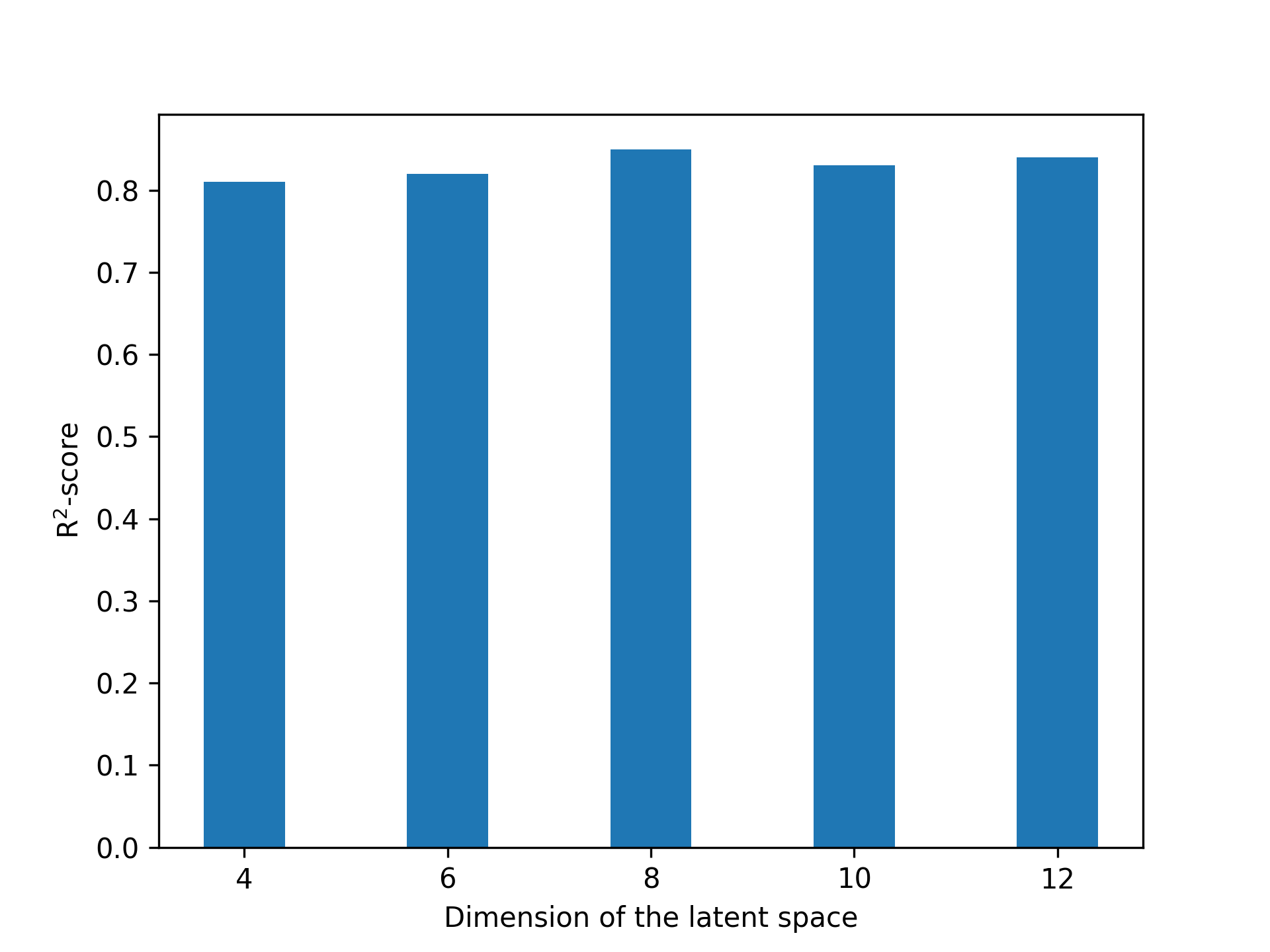}
    \caption{\(R^{2}\) plotted against various latent space dimensions. A latent space dimension of \(z=8\) yielded the most optimal fit for the MLP to the target.}
    \label{fig:latent}
\end{figure}

\begin{figure}[htbp]
    \centering
    \includegraphics[width=0.85\linewidth]{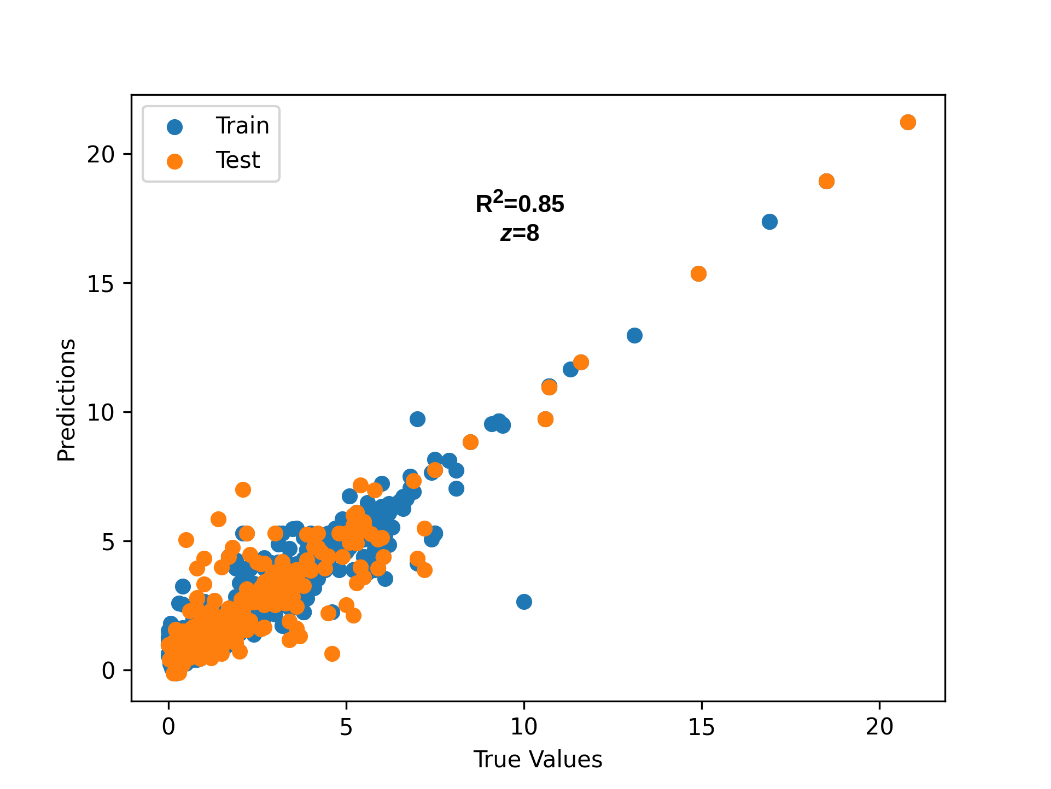}
    \caption{Pair plot comparing the true values against the predicted values (both test and train) for a latent space dimension of \(z=8\) and \(R^2=0.85\)}
    \label{fig:pair}
\end{figure}

Performance evaluation was done by computing the \(R^2\) value across different latent space dimensions. Our observations revealed that the \(R^2\) value exhibited an initial increase with the expansion of the latent space dimension, post which it fluctuated (Fig. \ref{fig:latent}) The peak \(R^2\) value (0.85) was obtained with a latent space dimension of \(z=8\) (Fig. \ref{fig:pair}).

This outcome suggests that a latent space dimension of 8 effectively encapsulates the vast majority of information inherent in the original feature space, facilitating accurate predictions for hydrogen storage capacity. Thus, for ($z< 8$) the latent space might be too small to capture all the relevant features of the input data. This can lead to underfitting, where the model fails to learn the underlying patterns effectively. The low dimensionality may force the autoencoder to compress the data so much that significant information is lost, making it difficult for the MLP to make accurate predictions. While for ($z>8$) the latent space might be too large, potentially leading to overfitting, where the model learns noise and details from the training data that do not generalize well to unseen data. A higher-dimensional latent space can represent the data more accurately, but it can also capture unnecessary or redundant information, which can degrade the performance of the MLP when it tries to generalize from the training data to new data.

\section{Comparison with few other supervised models}
\begin{figure}[htbp]
    \centering
    \includegraphics[width=1.1\linewidth]{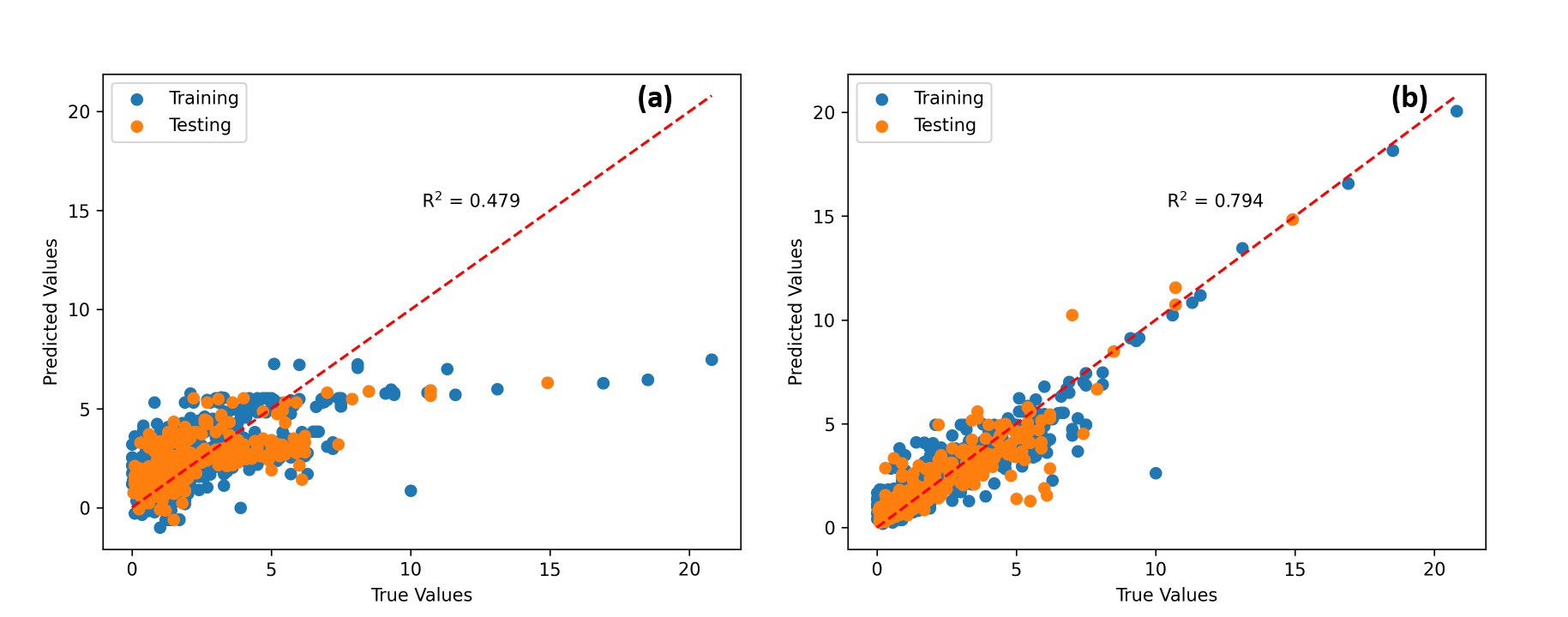}
    \caption{Performance of the Linear models (a) LASSO (b) GBR. Here the original 36 features are used.}
    \label{Linear}
\end{figure}

In this section, we compare the performance of two supervised models, Lasso regression and Gradient Boosting regression, with the previously mentioned hybrid approach using an Autoencoder combined with a Multilayer Perceptron (MLP) \cite{tibshirani1996regression,friedman2001greedy}. We evaluate the models based on their R$^2$ scores, which measure the proportion of the variance in the target variable that the models can explain.
The results are shown in Fig.\ref{Linear}.
The R$^2$ values obtained from the Lasso regression and Gradient Boosting regression are 0.48 and 0.8, respectively. These values indicate that both models have relatively lower predictive accuracy compared to the hybrid approach. The hybrid approach achieved an R$^2$ score of 0.85, outperforming both linear models. These results suggest that the hybrid approach, combining an Autoencoder with an MLP, is more effective in capturing the underlying patterns and predicting the target variable compared to the individual linear models.

\subsection{Identification of new hydrogen storage materials}
\begin{figure}[htbp]
	\centering
	\includegraphics[width=0.9\linewidth]{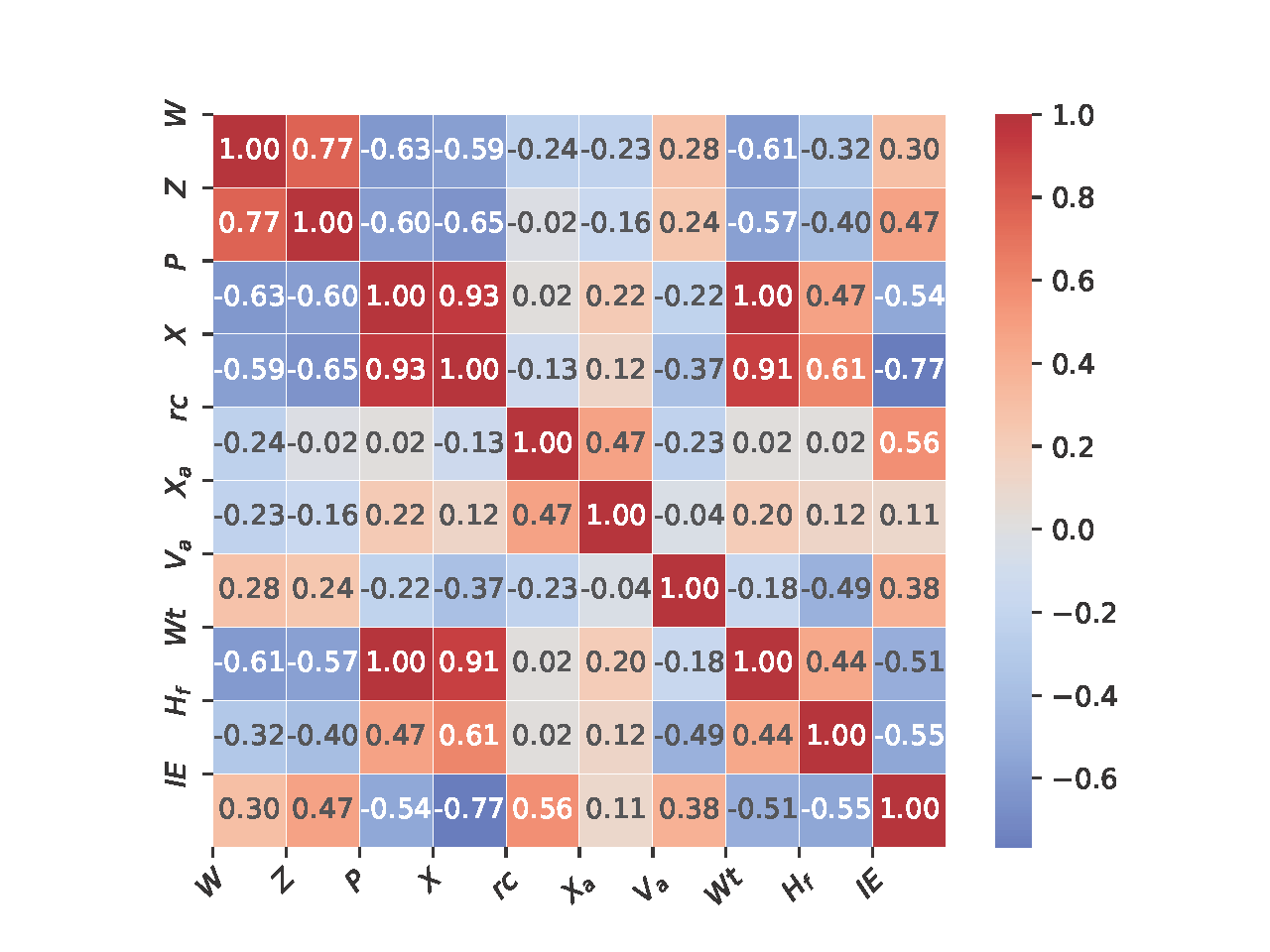}
	\caption{Correlation between different physically meaningful features and the target (Wt). Here 
  atomic weight (W), atomic number ($Z$), period ($P$), electronegativity ($\chi$), covalent radius ($r_c$), electron affinity ($\chi_a$), atomic volume ($V_a$), fusion heat ($H_f$), and ionization energy ($IE$) are considered.}
 \label{correlation}
\end{figure}

To find new materials for efficient hydrogen storage, we used the following approach: We calculated the Pearson correlation function for different physically meaningful features, including the target - the storage capacity (Wt), which we show in Fig.\ref{correlation} \cite{Chen_2021}. We take the weighted sum for all nine features and compute the correlation.
It can be seen that the period number and electronegativity have the highest correlation with the weight percentage. A significant Pearson correlation between the period number and electronegativity with the target hydrogen storage capacity, suggests their pivotal roles in influencing a material's ability to store hydrogen efficiently. The period number, reflecting the electron shell structure, directly impacts its hydrogen bonding capabilities. As the period number increases, so does the atomic radius. This can influence the hydrogen storage capacity because larger atoms can provide more interstitial sites for hydrogen atoms to occupy. On the other hand, electronegativity is a measure of the ability of an atom to attract and hold onto electrons. Elements with higher electronegativity tend to form more stable hydrides because they attract and bond more effectively with hydrogen atoms.  
Therefore, we first scanned the metal hydrides from the Materials Project with these two features. Next, we constructed a 36-dimensional feature space using the above materials, as before (using four statistical measures for each one). Then, we used pre-trained embeddings from the autoencoder so that the property (hydrogen weight in percentage) can be computed using the MLP. Finally, we performed DFT-based calculations to re-verify these values.

Using the above approach we identified two materials Al$_{11}$O$_{18}$ and V$_2$O$_5$. From our formation energy calculations, both materials are stable. The DFT calculated formation energies are respectively -3.34 eV/atom and  -2.12 eV/atom for Al$_{11}$O$_{18}$ and V$_2$O$_5$. From the DFT approach the hydrogen storage capacities were calculated using the hydrogen absorption energies~\cite{das2023computational} shown below,

\begin{equation}
	E_{abs} =\frac{1}{n}\times\left(E_{structure+nH}-\left( E_{structure}+\frac{n}{2}E_{H_2}\right)\right)
	\label{Eads}
\end{equation}

where E\textsubscript{structure+nH} is the energy of the hydrogenated structure, E\textsubscript{structure} and E\textsubscript{H\textsubscript{2}} are the energies of the pristine structure and isolated H\textsubscript{2}, respectively, and n is the number of hydrogen atoms involved in the absorption.

\begin{figure}[htbp]
    \includegraphics[width=1.1\linewidth]{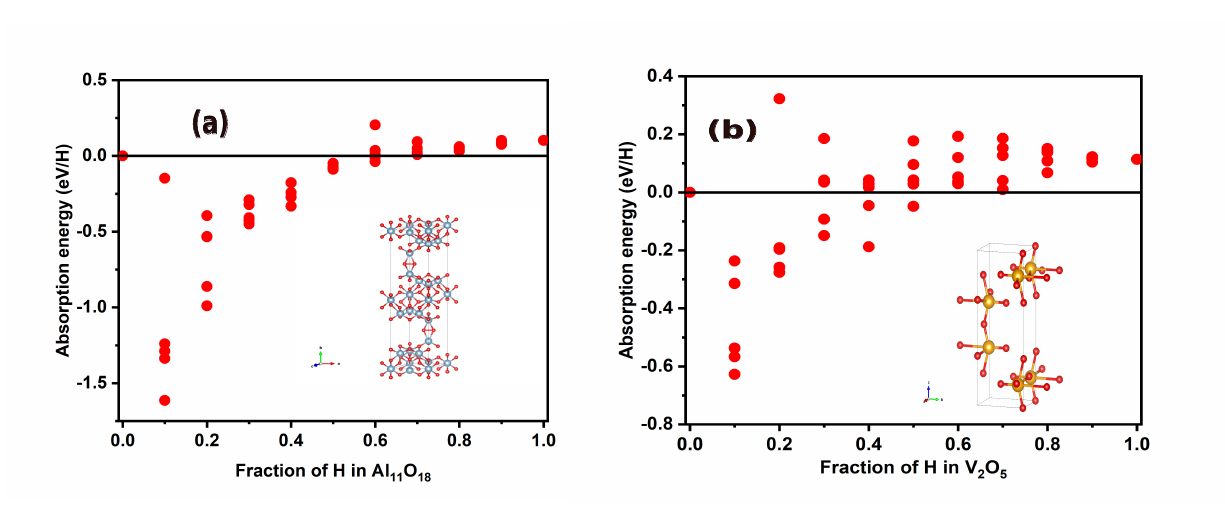}
    \caption{Absorption energy of hydrogen in Al$_{11}$O$_{18}$ (a) and V$_2$O$_5$ (b) at different H
concentration.}
    \label{capacity}
\end{figure}

To predict hydrogen storage capacity, all interstitial sites of Al$_{11}$O$_{18}$ and V$_2$O$_5$ are gradually filled with hydrogen, and the absorption energy is calculated \cite{das2023computational2}. Figure \ref{capacity} shows the change in absorption energy with increasing hydrogen concentration. The absorption energy increases as the hydrogen fraction increases. The maximum hydrogen fraction at which absorption energy remains negative is used to predict hydrogen storage capacity. The predicted hydrogen storage capacities from the DFT calculations are respectively 4.61\% and 3.83\% while from the MLP model, we obtain values respectively 5.98\% and 4.35\%.

\begin{figure}
    \centering
    \includegraphics[width=1.0\linewidth]{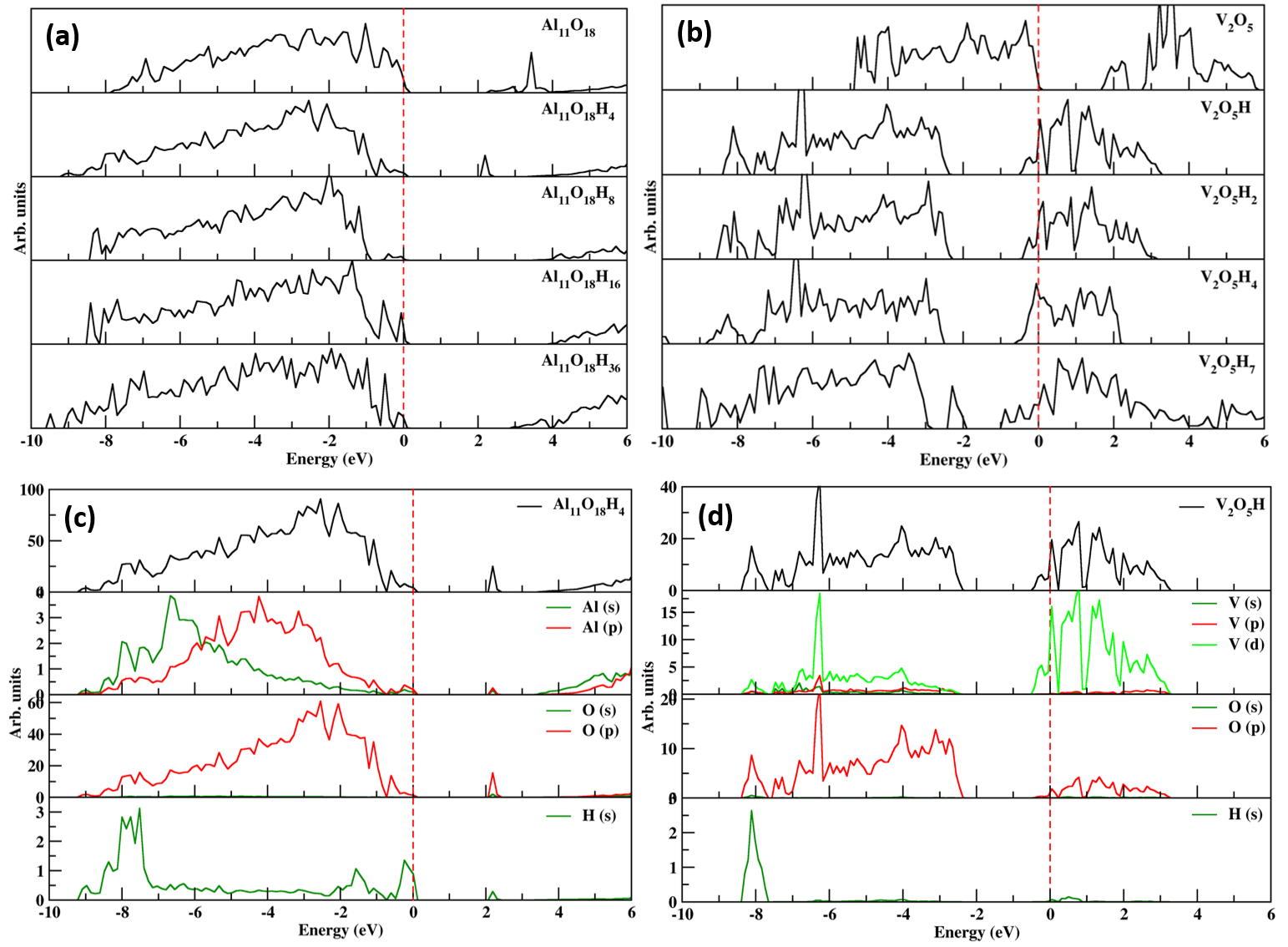}
    \caption{The total (top panel) and partial DOS (down panel) for the predicted materials Al$_{11}$O$_{18}$ V$_2$O$_5$ and their hydrides. The Fermi level is set at zero energy and marked by the vertical lines.}
    \label{DOS}
\end{figure}

In Fig.\ref{DOS}, we show the density of states (DOS) for the machine learning predicted materials and their hydrides for different hydrogen loadings. It was found that Al$_{11}$O$_{18}$ shows a more stable behavior when loaded with hydrogen compared to V$_2$O$_5$ because V$_2 $O$_5$ relatively DOS at the Fermi energy for all the hydrogen loading. For both materials, H-s strongly hybridizes with the O-p states as can be seen from the projected DOS of Al$_{11}$O$_{18}$H$_4$ and V$_2 $O$_5$H shown in the down panel of the above figure. In the case of V$_2 $O$_5$H, the anti-bonding states formed between V-d and H-s are also filled and therefore the system shows less stability. 
\subsection{Use of Partial Weight-Transfer and Freezing Encoder Weights}
So far, we have discussed the prediction of hydrogen weight percentage by directly feeding the latent features ($z$) to the MLP without considering the trained weights of the autoencoder. One might be interested to know what if along with the latent features the pre-trained weights from the autoencoder are also transferred in the downstream MLP task. In the supervised learning phase, we explored a hybrid approach that integrates weights from the trained autoencoder into the MLP for predictive modeling. 

The autoencoder follows the architecture of \texttt{dim\_in} $\rightarrow$ 256 $\rightarrow$ \texttt{dim\_latent} $\rightarrow$ 256 $\rightarrow$ \texttt{dim\_in}, encoding and then decoding the information. \texttt{dim\_in} is the original feature dimension (36), while \texttt{dim\_latent} is the dimension of the latent space. In contrast, the original MLP architecture is \texttt{dim\_latent} $\rightarrow$ 512 $\rightarrow$ 256 $\rightarrow$ 16 $\rightarrow$ 1, a straightforward feedforward network without the encoding-decoding mechanism.

To facilitate transfer learning and leverage the learned representations from the autoencoder, we adjusted the first two layers of the MLP to align with the autoencoder's encoder structure. The modified MLP structure is \texttt{dim\_latent} $\rightarrow$ 256 $\rightarrow$ \texttt{dim\_latent} (8) $\rightarrow$ 512 $\rightarrow$ 256 $\rightarrow$ 16 $\rightarrow$ 1. This structure begins by expanding \texttt{dim\_latent} to 256, then contracts back to \texttt{dim\_latent} (8), before proceeding through the remaining MLP layers.

\begin{figure}
    \centering
    \includegraphics[width=0.75\linewidth]{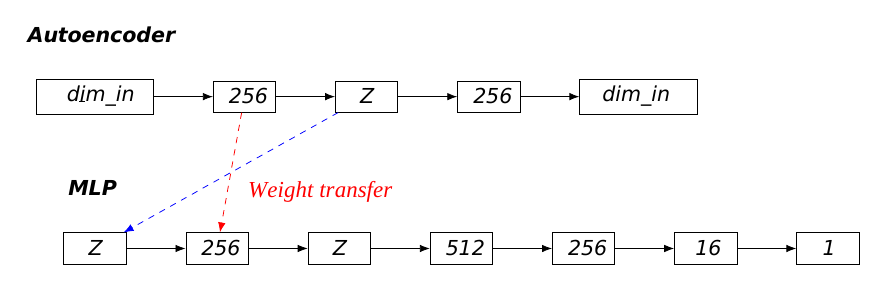}
    \caption{Schematic representation of partial weight transfer from the Autoencoder to the MLP. The Autoencoder encodes the input dimension into a latent space ($z$) and then decodes back to the original dimension. In the MLP, the first layer expands $z$ to 256, contracts back to $z$, and then proceeds through additional layers to reach the final single output (storage capacity). The red dashed line indicates the transfer of weights from the Autoencoder's encoder to the initial layer of the MLP.}
    \label{fig:transfer}
\end{figure}

As depicted in Fig.\ref{fig:transfer}, the process of weight transfer involves the transfer of weights from the encoder part of the autoencoder, specifically the weights leading into the latent space $z$, which are copied to the corresponding layer of MLP.

In addition, in order to safeguard the capability of the autoencoder to extract features, we maintained the weights of the transferred encoder layers in a frozen state throughout the training of the MLP. This was accomplished by ensuring that their weights remain unaltered during the backpropagation phase. By adopting this approach, the MLP is able to benefit from the pre-trained representations in the autoencoder while also acquiring the ability to learn features specific to the given task in subsequent layers. The amalgamation of weight transfer and selective weight freezing constitutes an innovative method in our modeling process. However, despite the presence of some improvements, it cannot be deemed as remarkable. Keeping the remaining parameters unchanged, we obtained an R-squared value of 0.86 for z=8, which represents a modest improvement when compared to our previous scenario where only the latent features were transferred to the MLP. Thus this study highlights the complexities and limitations of applying transfer learning techniques in the context of materials science. One possibility in this particular case off-course could be that the training of the MLP with fixed encoder weights may give rise to optimization difficulties, wherein the propagation of gradient updates across the network may not be efficient due to the unchangeable nature of specific weights. The other possibility could be that the initial task for which the autoencoder was trained may not align well with the new task in terms of weight, meaning that the weights learned may not be optimal for predicting the hydrogen weight percentage even though features were useful.

\section{Application to the Hydrogen Storage Materials Generated  Using a Large Language Model}
To show further the predictive capability of our approach, we apply it further to a set of materials that we generate using a large language model (LLM). Leveraging the power of GPT-2 (Generative Pre-trained Transformer version 2), a model renowned for its proficiency in natural language processing, we have generated a few new hydrogen storage materials. Unlike the resource-intensive traditional methods such as Generative Adversarial Networks (GANs) or diffusion models, which have limited testing in materials science, our approach utilizes a fine-tuned GPT-2 model, offering a more quick and targeted pathway to innovation. The details of the methodology are explained as detailed by Fu \textit{et. al}~\cite{fu2023material} which leverages the advanced capabilities of LLMs originally developed for natural language processing, adapting them to the intricate language of material compositions.
The fine-tuning process involves adapting a pre-trained GPT-2 model to the specific nuances of material science. This adaptation is facilitated by a custom vocabulary. which includes a comprehensive list of chemical elements and special tokens. This specialized vocabulary is crucial for accurately representing and processing material compositions. Our training process fine-tunes the model with a dataset specifically curated for hydrogen storage materials, allowing the model to learn and predict the complex patterns of material composition. The cornerstone of this methodology is the training of these models on comprehensive databases of known material compositions, including the ICSD (Inorganic Crystal Structure Database), OQMD (Open Quantum Materials Database), and Materials Project. This substantial assimilation of existing material data enables LLMs such as GPT-2 to understand and recreate the complex patterns and laws that govern material compositions. Through this method, the models can produce new, chemically realistic material compositions with high promise for hydrogen storage applications. 
In an extension to the work by Fu et al.~\cite{fu2023material} on materials generation, we take a slightly modified approach where not only the chemical composition of materials but also a specific property: their hydrogen storage capacity is also considered to generate the chemical labels.
\begin{figure}
    \centering
    \includegraphics[width=0.8\linewidth]{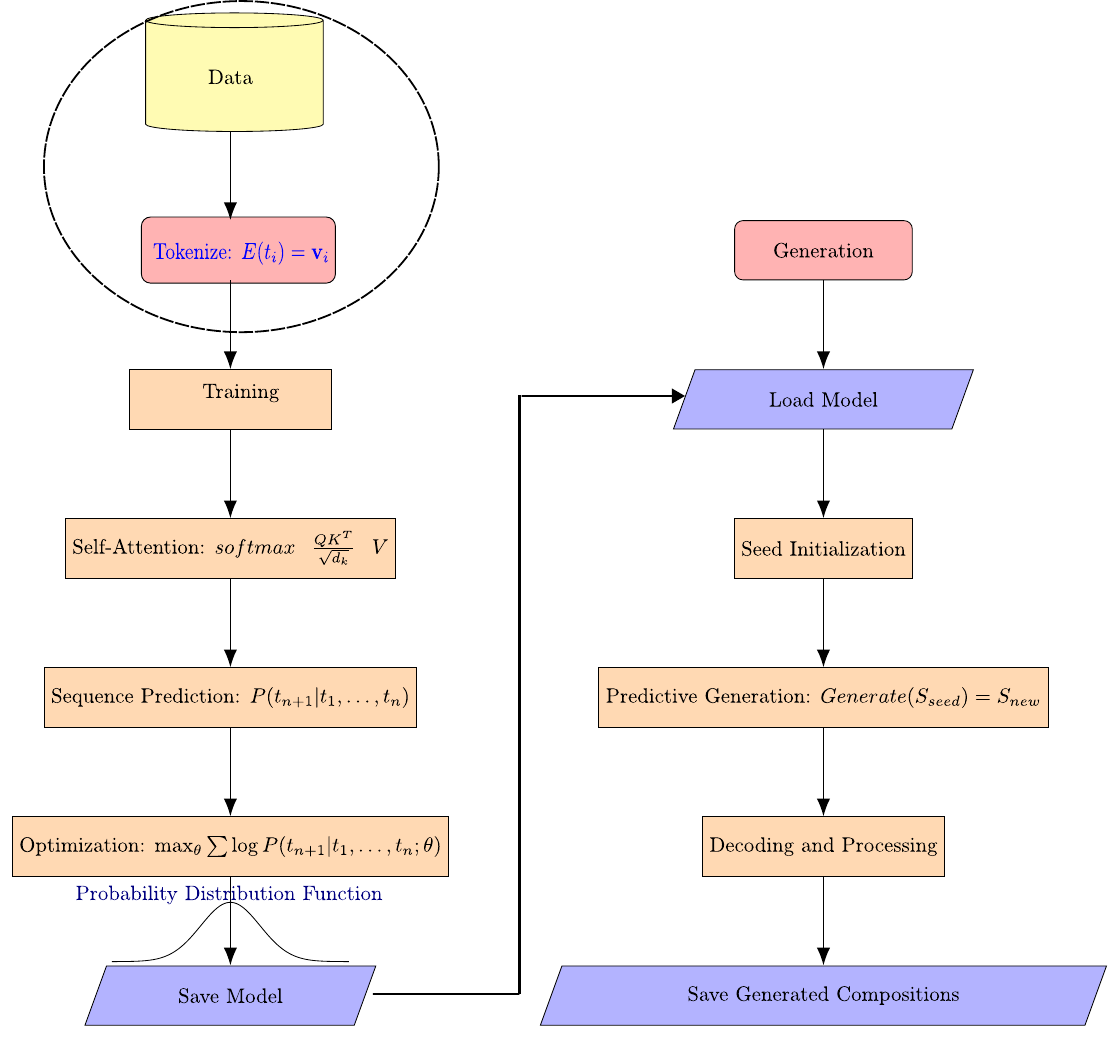}
    \caption{Flowchart depicting the training and generation processes for a GPT-2 language model. The left column (Training Phase) outlines the steps from initializing parameters to saving the trained model. The right column (Generation Phase) demonstrates the sequence generation process using the trained model.}
    \label{GP2-training}
\end{figure}
As a result, the training of the model emphasizes the detailed relationship between a material's composition and its storage capacity. Such integration enhances the model's capacity to predict materials more accurately. The model's configuration is tailored for this task, with an embedding size of 180, 2 attention heads, and 6 layers, making for a robust yet effective learning structure. Furthermore, the model's vocabulary has been expanded to include numerical values, facilitating the inclusion of hydrogen storage capacity data. The GPT-2 transformer model training for generating new hydrogen storage materials was conducted using the Huggingface Transformers library~\cite{huggingface}.

The training and generation of materials using GPT-2 transformer model is shown in the Fig.\ref{GP2-training}.
Starting from a sequence of input tokens (comprising chemical symbols and hydrogen storage capacity) the model first computes their embeddings. The application of learned weight matrices is subsequently utilized to deduce the conventional queries, keys, and values matrices \(Q\), \(K\), and \(V\), respectively. These matrices play a crucial role in the computation of self-attention scores, as expressed by $\text{softmax}\left(\frac{QK^T}{\sqrt{d_k}}\right)V$~\cite{vaswani2017attention}, where \( d_k \) represents the dimensionality of the key vectors. The evaluation of these scores dictates the level of attention that each token in the sequence should allocate to every other token, encompassing both local and global dependencies present in the sequence. During the training phase, the objective is to fine-tune a model's parameters, say denoted by \( \theta \), to accurately predict material compositions from a given sequence of elements. In this process, the model is presented with sequences of tokens \( [t_1, t_2, \ldots, t_n] \), each representing either a unique element within a material composition or a numerical value representing the hydrogen storage capacity. The model learns by adjusting \( \theta \) to maximize the log-likelihood of correctly predicting the next token \( t_{n+1} \) in the sequence, formalized as \( \max_{\theta} \sum \log P(t_{n+1} | t_1, \ldots, t_n; \theta) \). Here, \( P(t_{n+1} | t_1, \ldots, t_n) \) represents the conditional probability of getting the next token  \( t_{n+1} \) given the preceding sequence \( [t_1, \ldots, t_n] \). By enabling every token to dynamically affect the prediction of the subsequent token based on its context and relevance to the remainder of the sequence, the self-attention mechanism in the model helps with this learning process.

During generation, the trained model is loaded to predict new material compositions. Starting with a random seed \( S_{\text{rand}} \), which may consist of one or more element tokens and a random value of the storage capacity, the model iteratively generates the next element in the sequence by sampling from the probability distribution \( P(t_{n+1} | t_1, \ldots, t_n; \theta) \) it learned during training. This process continues, adding each newly predicted element to the growing composition, until a termination condition is met, such as reaching a predefined sequence length or generating a special end-of-sequence token. The generation process is informed by the patterns and rules learned during training (or precisely the grammar of the chemistry), enabling the exploration of new and potentially viable material compositions that were not present in the original training dataset.

The model leverages a large corpus of known material compositions, learning intricate relationships and dependencies between different elements. By incorporating these learned patterns, the model can generate sequences that adhere to known chemical rules and potentially suggest novel compositions that exhibit desirable properties. This approach significantly accelerates the discovery process by providing a vast space of candidate materials for further computational or experimental validation.

To ensure that the generated compositions are meaningful, a few steps are further used. Initially, the sequences are cleaned to remove special tokens and ensure they conform to valid chemical notation. Next, they are filtered based on predefined criteria, such as limiting the number of unique elements and the total atom count, to focus on manageable and realistic compositions. In our generation process, we set the filtering criteria to ensure that the generated material compositions do not have more than 8 unique elements and the total number of atoms in any generated composition does not exceed 30. Furthermore, such generation and post-processing steps combined lead to the generation of sequences focusing on chemical compositions without storage capacities. Therefore, we generate only chemical compositions not the storage capacities though the training data includes such capacities. 

Using this methodology, we have successfully generated 225 new materials with potential applications in hydrogen storage.

\begin{table}[H]
\centering
\begin{tabular}{|l|l|l|l|l|l|l|l|l|}
\hline 
\textbf{Materials} & \textbf{No of H} & \textbf{FE (eV)} & \textbf{$\mu_H^C$} & \textbf{$\mu_H^X$} & \textbf{Wt\% (DFT)} & \textbf{Wt\% (ML)} & \textbf{E$_h$} & \textbf{Expt.} \\
\hline 
$\mathrm{TiAlN_2}$ & 3 & 1.06 & -4.8 & -3.3 & 2.86 & 4.6& 0.21 & no\\
\hline 
$\mathrm{V}_2\mathrm{H}_2$ & 2+3 & -0.2 & -3.58 & -3.3 & 2.83 & 3.24 & 0.29 & no \\
\hline 
$\mathrm{Zr}_2\mathrm{Ti}\mathrm{Al}$ & 1 & -0.34 & -3.72 & -3.3 & 0.4 & 1.1 & 0.13 & yes \\
\hline 
$\mathrm{MgC}$ & 2 & 0.37 & -3.0 & -3.3 & 5.25 & 4.65 & 1.25 & no \\
\hline 
$\mathrm{NLi}$ & 3 & -1.67 & -5.05 & -3.3 & 6.73 & 4.5 & 1.34 & yes \\
\hline
\end{tabular}
\caption{Examples of hydrogen storage materials identified by GPT-2 model. From the Materials project, we report the structures with the lowest energy above Hull E$_h$ (eV), whether they are experimentally observed or not. From the hybrid method mentioned above (\text{Wt\% (ML)}), we calculated their storage capacity as well and compared it with the corresponding DFT-values (\text{Wt\% (DFT)}). The last column indicates whether the material is experimentally synthesized (yes) or not (no). }
\end{table}

Here the critical hydrogen chemical potential, $\mu_H^C$ is defined as  $\mu_H^C=1 / n\left(E_{M A X+n H}-E_{M A X}\right)$, which is lower than the chemical potential $\mu_H$ of the hydrogen at gas phase at very low temperature (as our calculation corresponds to DFT temperature scale). "FE" stands for the formation energy.
From the above table, it is apparent that our model exhibits commendable performance and is quite reliable for the calculation of the hydrogen storage capacity. In the final table \ref{tab:final_table}, we present an additional five materials that have been predicted by GPT-2 and whose hydrogen weight percentage has been computed by our model. As the atomic configurations of these materials remain elusive, we employ the methodology expounded by Kusaba \textit{et al}~\cite{KUSABA2022111496}, which employs the approach of \textit{metric learning}, where the algorithm has been trained on a substantial amount of crystal structures that have already been identified to ascertain the isomorphism of crystal structures formed by two specified chemical compositions, yielding an accuracy of approximately 96.4\%. For a given composition with an unknown crystal structure, this approach automatically selects a collection of template crystals from a database of crystal structures that possess nearly identical stable structures. When we subject the compositions predicted by GPT-2 to this method, we obtain a collection of structures for each composition. We consider the one with the highest spacegroup and then perform the structural optimization followed by the calculation of the storage capacity using our hybrid-ML approach. The rightmost column of the table visually represents the crystal structure of each material. 

\begin{table}[H]
\centering
\begin{tabular}{|c|c|c|c|}
\hline
Material & \textbf{Wt\% (ML)} & \textbf{Spacegroup} & \textbf{Structure} \\ \hline
NMn$_2$Ti & 3.23 & 194 & \includegraphics[width=2 cm]{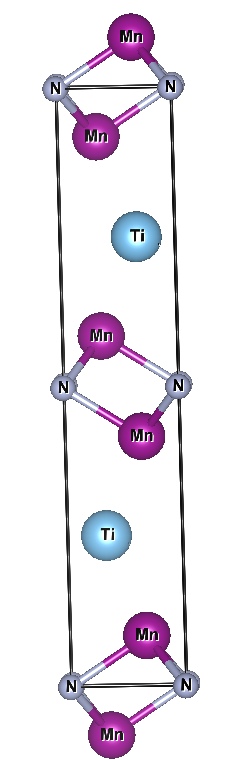} \\ \hline
MgCHF & 6.75 & 186 & \includegraphics[width=2 cm]{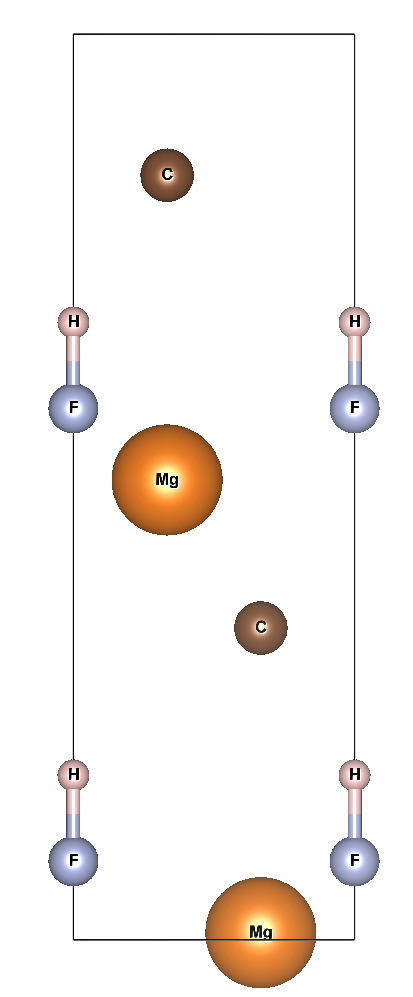} \\ \hline
CAlB & 6.24 & 194 & \includegraphics[width=2 cm]{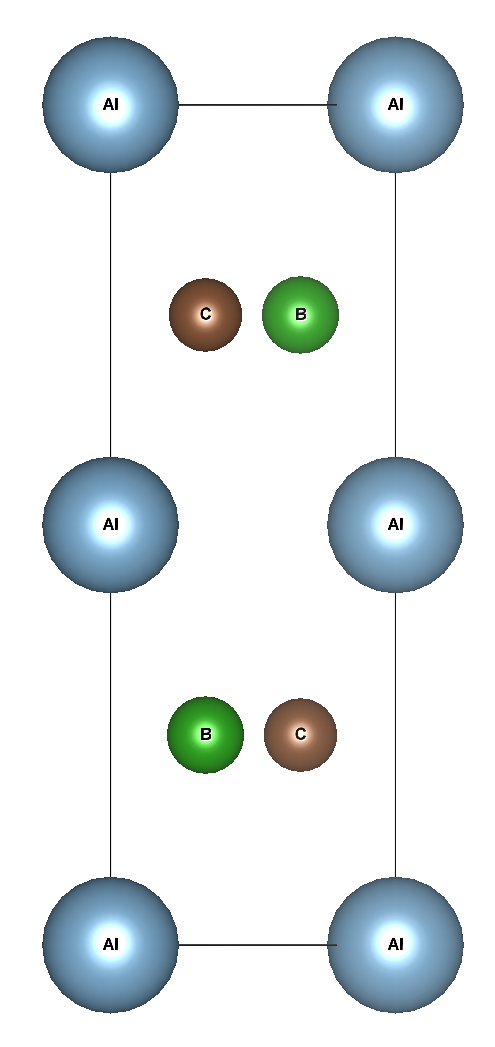} \\ \hline
MgCF & 5.69 & 129 & \includegraphics[width=2 cm]{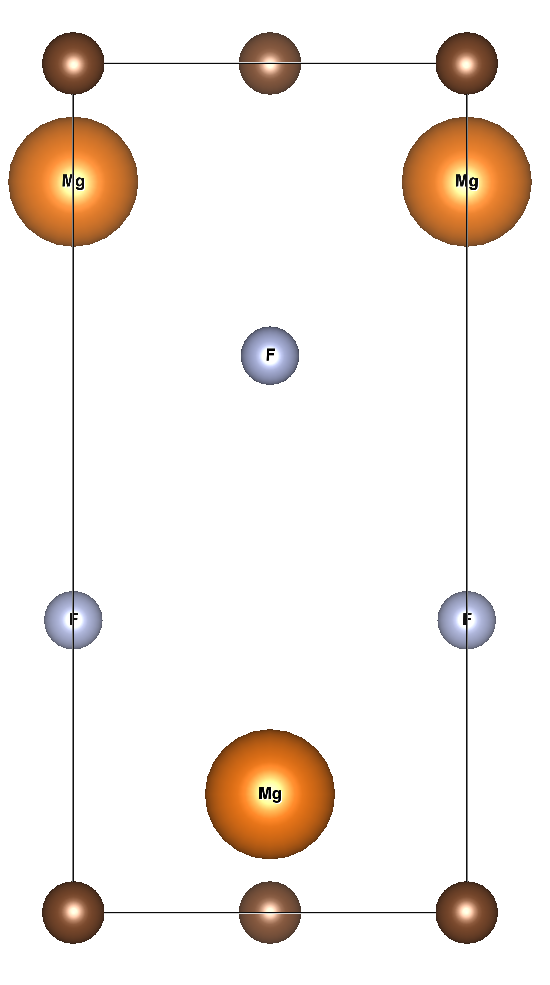} \\ \hline
MgMnVTi & 1.93 & 216& \includegraphics[width=3.3 cm]{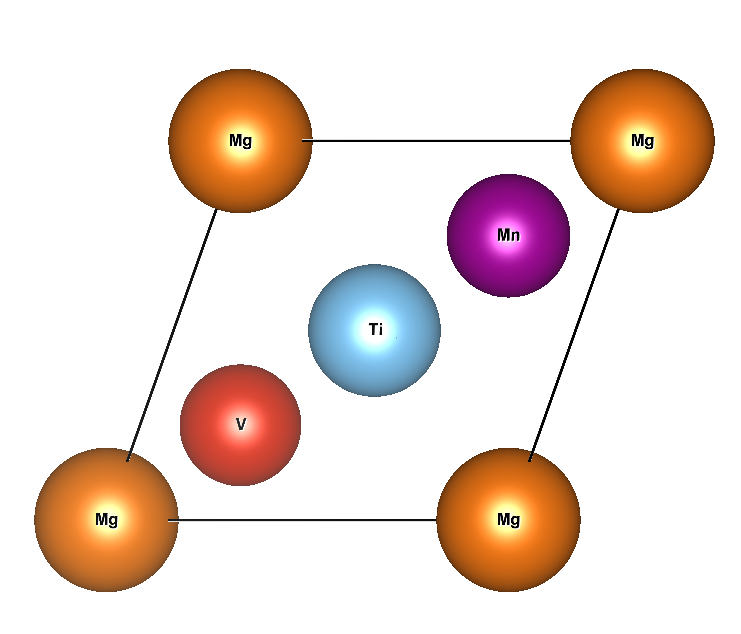} \\ \hline
\end{tabular}
\caption{Materials predicted by GPT-2 model and their storage capacities obtained using our hybrid ML approach. The spacegroup information was obtained via DFT optimization of the structure with the highest spacegroup predicted by the approach mentioned in Kusaba \textit{et al}~\cite{KUSABA2022111496} }
\label{tab:final_table}
\end{table}
It will be worth synthesizing experimentally these materials for enhanced hydrogen storage applications. In future work, we plan to explore the use of advanced models like GPT-3.5 which could be highly beneficial. GPT-3.5, with its sophisticated natural language understanding and generation capabilities, can potentially enhance the prediction and design of novel hydrogen storage materials. Its advanced algorithms and large knowledge base might lead to more accurate predictions and innovative approaches in material science, opening new directions for efficient and effective hydrogen storage solutions.
\section{Conclusion}
This study investigated the efficacy of a hybrid deep learning framework for predicting hydrogen storage capacities in metal hydrides, addressing the critical issue of limited experimental data. By leveraging an advanced autoencoder and elemental descriptors from Mendeleev software, we effectively tackled the challenge of high-dimensional feature spaces inherent to materials informatics, where data is often sparse and heterogeneously distributed. A deep Multi-Layer Perceptron (MLP) architecture with five layers served as the prediction model, utilizing latent representations generated by the autoencoder to capture crucial features from high-dimensional data. The framework was tested for a few materials that were selected as potential hydrogen storage materials from the features that are highly correlated with the target as well as for the materials that are generated by a large language model: GPT-2. In both cases, the predicted capacity by the ML approach matches quite well with the DFT calculations. We further discuss the issue of weight transfer from the Autoencoder to the MLP in addition to the latent features.

\section*{Acknowledgements}
This work was supported by the Korea Institute of Science and Technology (Grant number 2E31851), GKP (Global Knowledge Platform, Grant number 2V6760) project of the Ministry of Science, ICT and Future Planning.

\section{Conflict of interest}
The authors declare no conflict of interest.
\bibliography{He}

\providecommand{\latin}[1]{#1}
\makeatletter
\providecommand{\doi}
  {\begingroup\let\do\@makeother\dospecials
  \catcode`\{=1 \catcode`\}=2 \doi@aux}
\providecommand{\doi@aux}[1]{\endgroup\texttt{#1}}
\makeatother
\providecommand*\mcitethebibliography{\thebibliography}
\csname @ifundefined\endcsname{endmcitethebibliography}  {\let\endmcitethebibliography\endthebibliography}{}
\begin{mcitethebibliography}{66}
\providecommand*\natexlab[1]{#1}
\providecommand*\mciteSetBstSublistMode[1]{}
\providecommand*\mciteSetBstMaxWidthForm[2]{}
\providecommand*\mciteBstWouldAddEndPuncttrue
  {\def\EndOfBibitem{\unskip.}}
\providecommand*\mciteBstWouldAddEndPunctfalse
  {\let\EndOfBibitem\relax}
\providecommand*\mciteSetBstMidEndSepPunct[3]{}
\providecommand*\mciteSetBstSublistLabelBeginEnd[3]{}
\providecommand*\EndOfBibitem{}
\mciteSetBstSublistMode{f}
\mciteSetBstMaxWidthForm{subitem}{(\alph{mcitesubitemcount})}
\mciteSetBstSublistLabelBeginEnd
  {\mcitemaxwidthsubitemform\space}
  {\relax}
  {\relax}

\bibitem[Z{\"u}ttel(2003)]{zuttel2003materials}
Z{\"u}ttel,~A. Materials for hydrogen storage. \emph{Materials today} \textbf{2003}, \emph{6}, 24--33\relax
\mciteBstWouldAddEndPuncttrue
\mciteSetBstMidEndSepPunct{\mcitedefaultmidpunct}
{\mcitedefaultendpunct}{\mcitedefaultseppunct}\relax
\EndOfBibitem
\bibitem[Vezirolu and Barbir(1992)Vezirolu, and Barbir]{Wonder}
Vezirolu,~T.; Barbir,~F. Hydrogen: the wonder fuel. \emph{International Journal of Hydrogen Energy} \textbf{1992}, \emph{17}, 391--404\relax
\mciteBstWouldAddEndPuncttrue
\mciteSetBstMidEndSepPunct{\mcitedefaultmidpunct}
{\mcitedefaultendpunct}{\mcitedefaultseppunct}\relax
\EndOfBibitem
\bibitem[Barth{\'e}l{\'e}my \latin{et~al.}(2017)Barth{\'e}l{\'e}my, Weber, and Barbier]{barthelemy2017hydrogen}
Barth{\'e}l{\'e}my,~H.; Weber,~M.; Barbier,~F. Hydrogen storage: Recent improvements and industrial perspectives. \emph{International Journal of Hydrogen Energy} \textbf{2017}, \emph{42}, 7254--7262\relax
\mciteBstWouldAddEndPuncttrue
\mciteSetBstMidEndSepPunct{\mcitedefaultmidpunct}
{\mcitedefaultendpunct}{\mcitedefaultseppunct}\relax
\EndOfBibitem
\bibitem[Oesterreicher(1981)]{oesterreicher1981hydrides}
Oesterreicher,~H. Hydrides of intermetallic compounds. \emph{Applied physics} \textbf{1981}, \emph{24}, 169--186\relax
\mciteBstWouldAddEndPuncttrue
\mciteSetBstMidEndSepPunct{\mcitedefaultmidpunct}
{\mcitedefaultendpunct}{\mcitedefaultseppunct}\relax
\EndOfBibitem
\bibitem[Lee and Lee(2005)Lee, and Lee]{lee2005review}
Lee,~B.-K.; Lee,~K.-S. A review on hydrogen storage in metal hydrides. \emph{International Journal of Hydrogen Energy} \textbf{2005}, \emph{30}, 947--958\relax
\mciteBstWouldAddEndPuncttrue
\mciteSetBstMidEndSepPunct{\mcitedefaultmidpunct}
{\mcitedefaultendpunct}{\mcitedefaultseppunct}\relax
\EndOfBibitem
\bibitem[Churchard \latin{et~al.}(2011)Churchard, Banach, Borgschulte, Caputo, Caputo, Chen, Clary, Fijalkowski, Geerlings, Geerlings, Genova, Grochala, Jaroń, Juanes-Marcos, Kasemo, Kroes, Ljubić, Naujoks, Nørskov, Nørskov, Olsen, Pendolino, Remhof, Románszki, Tekin, Tekin, Vegge, Zäch, and Züttel]{Norskov}
Churchard,~A.~J. \latin{et~al.}  A multifaceted approach to hydrogen storage. \emph{Physical Chemistry Chemical Physics} \textbf{2011}, \emph{13}, 16955--16972\relax
\mciteBstWouldAddEndPuncttrue
\mciteSetBstMidEndSepPunct{\mcitedefaultmidpunct}
{\mcitedefaultendpunct}{\mcitedefaultseppunct}\relax
\EndOfBibitem
\bibitem[Lai \latin{et~al.}(2015)Lai, Paskevicius, Sheppard, Buckley, Thornton, Hill, Gu, Mao, Huang, Liu, \latin{et~al.} others]{lai2015hydrogen}
Lai,~Q.; Paskevicius,~M.; Sheppard,~D.~A.; Buckley,~C.~E.; Thornton,~A.~W.; Hill,~M.~R.; Gu,~Q.; Mao,~J.; Huang,~Z.; Liu,~H.~K.; others Hydrogen storage materials for mobile and stationary applications: current state of the art. \emph{ChemSusChem} \textbf{2015}, \emph{8}, 2789--2825\relax
\mciteBstWouldAddEndPuncttrue
\mciteSetBstMidEndSepPunct{\mcitedefaultmidpunct}
{\mcitedefaultendpunct}{\mcitedefaultseppunct}\relax
\EndOfBibitem
\bibitem[Astle(1974)]{astle1974crc}
Astle,~M.~J. \emph{CRC Han book}; CRC Press, 1974\relax
\mciteBstWouldAddEndPuncttrue
\mciteSetBstMidEndSepPunct{\mcitedefaultmidpunct}
{\mcitedefaultendpunct}{\mcitedefaultseppunct}\relax
\EndOfBibitem
\bibitem[Gray \latin{et~al.}(2011)Gray, Blach, Pitt, and Cookson]{GRAY20111630}
Gray,~E.; Blach,~T.; Pitt,~M.; Cookson,~D. Mechanism of the $\alpha$-to-$\beta$ phase transformation in the LaNi5–H2 system. \emph{Journal of Alloys and Compounds} \textbf{2011}, \emph{509}, 1630--1635\relax
\mciteBstWouldAddEndPuncttrue
\mciteSetBstMidEndSepPunct{\mcitedefaultmidpunct}
{\mcitedefaultendpunct}{\mcitedefaultseppunct}\relax
\EndOfBibitem
\bibitem[David(2005)]{DAVID2005169}
David,~E. An overview of advanced materials for hydrogen storage. \emph{Journal of Materials Processing Technology} \textbf{2005}, \emph{162-163}, 169--177, AMPT/AMME05\relax
\mciteBstWouldAddEndPuncttrue
\mciteSetBstMidEndSepPunct{\mcitedefaultmidpunct}
{\mcitedefaultendpunct}{\mcitedefaultseppunct}\relax
\EndOfBibitem
\bibitem[Verma \latin{et~al.}(2023)Verma, Mishra, Mukhopadhyay, and Yadav]{VERMA2023}
Verma,~S.~K.; Mishra,~S.~S.; Mukhopadhyay,~N.~K.; Yadav,~T.~P. Superior catalytic action of high-entropy alloy on hydrogen sorption properties of MgH2. \emph{International Journal of Hydrogen Energy} \textbf{2023}, \relax
\mciteBstWouldAddEndPunctfalse
\mciteSetBstMidEndSepPunct{\mcitedefaultmidpunct}
{}{\mcitedefaultseppunct}\relax
\EndOfBibitem
\bibitem[Das \latin{et~al.}(2023)Das, Lee, Lee, and Bhattacharjee]{das2023computational}
Das,~P.; Lee,~Y.-S.; Lee,~S.-C.; Bhattacharjee,~S. Computational design of a new palladium alloy with efficient hydrogen storage capacity and hydrogenation-dehydrogenation kinetics. \emph{International Journal of Hydrogen Energy} \textbf{2023}, \emph{48}, 18795--18803\relax
\mciteBstWouldAddEndPuncttrue
\mciteSetBstMidEndSepPunct{\mcitedefaultmidpunct}
{\mcitedefaultendpunct}{\mcitedefaultseppunct}\relax
\EndOfBibitem
\bibitem[Li \latin{et~al.}(2011)Li, Peng, Zhou, and Wan]{LI201114512}
Li,~C.; Peng,~P.; Zhou,~D.; Wan,~L. Research progress in LiBH4 for hydrogen storage: A review. \emph{International Journal of Hydrogen Energy} \textbf{2011}, \emph{36}, 14512--14526\relax
\mciteBstWouldAddEndPuncttrue
\mciteSetBstMidEndSepPunct{\mcitedefaultmidpunct}
{\mcitedefaultendpunct}{\mcitedefaultseppunct}\relax
\EndOfBibitem
\bibitem[Ali and Ismail(2021)Ali, and Ismail]{ALI20211111}
Ali,~N.; Ismail,~M. Advanced hydrogen storage of the Mg–Na–Al system: A review. \emph{Journal of Magnesium and Alloys} \textbf{2021}, \emph{9}, 1111--1122\relax
\mciteBstWouldAddEndPuncttrue
\mciteSetBstMidEndSepPunct{\mcitedefaultmidpunct}
{\mcitedefaultendpunct}{\mcitedefaultseppunct}\relax
\EndOfBibitem
\bibitem[Barthelemy \latin{et~al.}(2017)Barthelemy, Weber, and Barbier]{BARTHELEMY20177254}
Barthelemy,~H.; Weber,~M.; Barbier,~F. Hydrogen storage: Recent improvements and industrial perspectives. \emph{International Journal of Hydrogen Energy} \textbf{2017}, \emph{42}, 7254--7262\relax
\mciteBstWouldAddEndPuncttrue
\mciteSetBstMidEndSepPunct{\mcitedefaultmidpunct}
{\mcitedefaultendpunct}{\mcitedefaultseppunct}\relax
\EndOfBibitem
\bibitem[Westlake(1983)]{WESTLAKE19831}
Westlake,~D. Hydrides of intermetallic compounds: A review of stabilities, stoichiometries and preferred hydrogen sites. \emph{Journal of the Less Common Metals} \textbf{1983}, \emph{91}, 1--20\relax
\mciteBstWouldAddEndPuncttrue
\mciteSetBstMidEndSepPunct{\mcitedefaultmidpunct}
{\mcitedefaultendpunct}{\mcitedefaultseppunct}\relax
\EndOfBibitem
\bibitem[Lototskyy \latin{et~al.}(2017)Lototskyy, Tolj, Pickering, Sita, Barbir, and Yartys]{LOTOTSKYY20173}
Lototskyy,~M.~V.; Tolj,~I.; Pickering,~L.; Sita,~C.; Barbir,~F.; Yartys,~V. The use of metal hydrides in fuel cell applications. \emph{Progress in Natural Science: Materials International} \textbf{2017}, \emph{27}, 3--20\relax
\mciteBstWouldAddEndPuncttrue
\mciteSetBstMidEndSepPunct{\mcitedefaultmidpunct}
{\mcitedefaultendpunct}{\mcitedefaultseppunct}\relax
\EndOfBibitem
\bibitem[Lototskyy \latin{et~al.}(2014)Lototskyy, Yartys, Pollet, and Bowman]{LOTOTSKYY20145818}
Lototskyy,~M.; Yartys,~V.; Pollet,~B.; Bowman,~R. Metal hydride hydrogen compressors: A review. \emph{International Journal of Hydrogen Energy} \textbf{2014}, \emph{39}, 5818--5851\relax
\mciteBstWouldAddEndPuncttrue
\mciteSetBstMidEndSepPunct{\mcitedefaultmidpunct}
{\mcitedefaultendpunct}{\mcitedefaultseppunct}\relax
\EndOfBibitem
\bibitem[Tarasov \latin{et~al.}(2018)Tarasov, Bocharnikov, Yanenko, Fursikov, and Lototskyy]{TARASOV20184415}
Tarasov,~B.~P.; Bocharnikov,~M.~S.; Yanenko,~Y.~B.; Fursikov,~P.~V.; Lototskyy,~M.~V. Cycling stability of RNi5 (R=La, La+Ce) hydrides during the operation of metal hydride hydrogen compressor. \emph{International Journal of Hydrogen Energy} \textbf{2018}, \emph{43}, 4415--4427\relax
\mciteBstWouldAddEndPuncttrue
\mciteSetBstMidEndSepPunct{\mcitedefaultmidpunct}
{\mcitedefaultendpunct}{\mcitedefaultseppunct}\relax
\EndOfBibitem
\bibitem[Yao \latin{et~al.}(2018)Yao, Liu, Xiao, Wang, Jiang, and Chen]{YAO2018524}
Yao,~Z.; Liu,~L.; Xiao,~X.; Wang,~C.; Jiang,~L.; Chen,~L. Effect of rare earth doping on the hydrogen storage performance of Ti1.02Cr1.1Mn0.3Fe0.6 alloy for hybrid hydrogen storage application. \emph{Journal of Alloys and Compounds} \textbf{2018}, \emph{731}, 524--530\relax
\mciteBstWouldAddEndPuncttrue
\mciteSetBstMidEndSepPunct{\mcitedefaultmidpunct}
{\mcitedefaultendpunct}{\mcitedefaultseppunct}\relax
\EndOfBibitem
\bibitem[Sandrock(1999)]{SANDROCK1999877}
Sandrock,~G. A panoramic overview of hydrogen storage alloys from a gas reaction point of view. \emph{Journal of Alloys and Compounds} \textbf{1999}, \emph{293-295}, 877--888\relax
\mciteBstWouldAddEndPuncttrue
\mciteSetBstMidEndSepPunct{\mcitedefaultmidpunct}
{\mcitedefaultendpunct}{\mcitedefaultseppunct}\relax
\EndOfBibitem
\bibitem[Fashu \latin{et~al.}(2020)Fashu, Lototskyy, Davids, Pickering, Linkov, Tai, Renheng, Fangming, Fursikov, and Tarasov]{FASHU2020108295}
Fashu,~S.; Lototskyy,~M.; Davids,~M.~W.; Pickering,~L.; Linkov,~V.; Tai,~S.; Renheng,~T.; Fangming,~X.; Fursikov,~P.~V.; Tarasov,~B.~P. A review on crucibles for induction melting of titanium alloys. \emph{Materials \& Design} \textbf{2020}, \emph{186}, 108295\relax
\mciteBstWouldAddEndPuncttrue
\mciteSetBstMidEndSepPunct{\mcitedefaultmidpunct}
{\mcitedefaultendpunct}{\mcitedefaultseppunct}\relax
\EndOfBibitem
\bibitem[Niaz \latin{et~al.}(2015)Niaz, Manzoor, and Pandith]{NIAZ2015457}
Niaz,~S.; Manzoor,~T.; Pandith,~A.~H. Hydrogen storage: Materials, methods and perspectives. \emph{Renewable and Sustainable Energy Reviews} \textbf{2015}, \emph{50}, 457--469\relax
\mciteBstWouldAddEndPuncttrue
\mciteSetBstMidEndSepPunct{\mcitedefaultmidpunct}
{\mcitedefaultendpunct}{\mcitedefaultseppunct}\relax
\EndOfBibitem
\bibitem[Wei \latin{et~al.}(2017)Wei, Lim, Tseng, and Chan]{WEI20171122}
Wei,~T.; Lim,~K.; Tseng,~Y.; Chan,~S. A review on the characterization of hydrogen in hydrogen storage materials. \emph{Renewable and Sustainable Energy Reviews} \textbf{2017}, \emph{79}, 1122--1133\relax
\mciteBstWouldAddEndPuncttrue
\mciteSetBstMidEndSepPunct{\mcitedefaultmidpunct}
{\mcitedefaultendpunct}{\mcitedefaultseppunct}\relax
\EndOfBibitem
\bibitem[Zotov \latin{et~al.}(2008)Zotov, Movlaev, Mitrokhin, and Verbetsky]{ZOTOV2008220}
Zotov,~T.; Movlaev,~E.; Mitrokhin,~S.; Verbetsky,~V. Interaction in (Ti,Sc)Fe2–H2 and (Zr,Sc)Fe2–H2 systems. \emph{Journal of Alloys and Compounds} \textbf{2008}, \emph{459}, 220--224\relax
\mciteBstWouldAddEndPuncttrue
\mciteSetBstMidEndSepPunct{\mcitedefaultmidpunct}
{\mcitedefaultendpunct}{\mcitedefaultseppunct}\relax
\EndOfBibitem
\bibitem[Chen \latin{et~al.}(2020)Chen, Zhang, and Zhou]{dd1}
Chen,~A.; Zhang,~X.; Zhou,~Z. Machine learning: accelerating materials development for energy storage and conversion. \emph{InfoMat} \textbf{2020}, \emph{2}, 553--576\relax
\mciteBstWouldAddEndPuncttrue
\mciteSetBstMidEndSepPunct{\mcitedefaultmidpunct}
{\mcitedefaultendpunct}{\mcitedefaultseppunct}\relax
\EndOfBibitem
\bibitem[Kalinin \latin{et~al.}(2021)Kalinin, Zhang, Valleti, Pyles, Baker, De~Yoreo, and Ziatdinov]{dd2}
Kalinin,~S.~V.; Zhang,~S.; Valleti,~M.; Pyles,~H.; Baker,~D.; De~Yoreo,~J.~J.; Ziatdinov,~M. Disentangling rotational dynamics and ordering transitions in a system of self-organizing protein nanorods via rotationally invariant latent representations. \emph{ACS nano} \textbf{2021}, \emph{15}, 6471--6480\relax
\mciteBstWouldAddEndPuncttrue
\mciteSetBstMidEndSepPunct{\mcitedefaultmidpunct}
{\mcitedefaultendpunct}{\mcitedefaultseppunct}\relax
\EndOfBibitem
\bibitem[Kadurin \latin{et~al.}(2017)Kadurin, Nikolenko, Khrabrov, Aliper, and Zhavoronkov]{dd3}
Kadurin,~A.; Nikolenko,~S.; Khrabrov,~K.; Aliper,~A.; Zhavoronkov,~A. druGAN: an advanced generative adversarial autoencoder model for de novo generation of new molecules with desired molecular properties in silico. \emph{Molecular pharmaceutics} \textbf{2017}, \emph{14}, 3098--3104\relax
\mciteBstWouldAddEndPuncttrue
\mciteSetBstMidEndSepPunct{\mcitedefaultmidpunct}
{\mcitedefaultendpunct}{\mcitedefaultseppunct}\relax
\EndOfBibitem
\bibitem[Sparks \latin{et~al.}(2016)Sparks, Gaultois, Oliynyk, Brgoch, and Meredig]{SPARKS201610}
Sparks,~T.~D.; Gaultois,~M.~W.; Oliynyk,~A.; Brgoch,~J.; Meredig,~B. Data mining our way to the next generation of thermoelectrics. \emph{Scripta Materialia} \textbf{2016}, \emph{111}, 10--15, Viewpoint Set No. 57: Contemporary Innovations for Thermoelectrics Research and Development\relax
\mciteBstWouldAddEndPuncttrue
\mciteSetBstMidEndSepPunct{\mcitedefaultmidpunct}
{\mcitedefaultendpunct}{\mcitedefaultseppunct}\relax
\EndOfBibitem
\bibitem[Rahnama \latin{et~al.}(2018)Rahnama, Clark, and Sridhar]{RAHNAMA2018169}
Rahnama,~A.; Clark,~S.; Sridhar,~S. Machine learning for predicting occurrence of interphase precipitation in HSLA steels. \emph{Computational Materials Science} \textbf{2018}, \emph{154}, 169--177\relax
\mciteBstWouldAddEndPuncttrue
\mciteSetBstMidEndSepPunct{\mcitedefaultmidpunct}
{\mcitedefaultendpunct}{\mcitedefaultseppunct}\relax
\EndOfBibitem
\bibitem[Ramprasad \latin{et~al.}(2017)Ramprasad, Batra, Pilania, Mannodi-Kanakkithodi, and Kim]{ramprasad2017machine}
Ramprasad,~R.; Batra,~R.; Pilania,~G.; Mannodi-Kanakkithodi,~A.; Kim,~C. Machine learning in materials informatics: recent applications and prospects. \emph{npj Computational Materials} \textbf{2017}, \emph{3}, 54\relax
\mciteBstWouldAddEndPuncttrue
\mciteSetBstMidEndSepPunct{\mcitedefaultmidpunct}
{\mcitedefaultendpunct}{\mcitedefaultseppunct}\relax
\EndOfBibitem
\bibitem[Louis \latin{et~al.}(2022)Louis, Siriwardane, Joshi, Omee, Kumar, and Hu]{louis2022accurate}
Louis,~S.-Y.; Siriwardane,~E. M.~D.; Joshi,~R.~P.; Omee,~S.~S.; Kumar,~N.; Hu,~J. Accurate prediction of voltage of battery electrode materials using attention-based graph neural networks. \emph{ACS Applied Materials \& Interfaces} \textbf{2022}, \emph{14}, 26587--26594\relax
\mciteBstWouldAddEndPuncttrue
\mciteSetBstMidEndSepPunct{\mcitedefaultmidpunct}
{\mcitedefaultendpunct}{\mcitedefaultseppunct}\relax
\EndOfBibitem
\bibitem[Bhattacharjee and Lee(2022)Bhattacharjee, and Lee]{bhattacharjee2022general}
Bhattacharjee,~S.; Lee,~S.-C. A general rule for predicting the magnetic moment of Cobalt-based Heusler compounds using compressed sensing and density functional theory. \emph{Journal of Magnetism and Magnetic Materials} \textbf{2022}, \emph{563}, 169818\relax
\mciteBstWouldAddEndPuncttrue
\mciteSetBstMidEndSepPunct{\mcitedefaultmidpunct}
{\mcitedefaultendpunct}{\mcitedefaultseppunct}\relax
\EndOfBibitem
\bibitem[Ahmed and Siegel(2021)Ahmed, and Siegel]{ahmed2021predicting}
Ahmed,~A.; Siegel,~D.~J. Predicting hydrogen storage in MOFs via machine learning. \emph{Patterns} \textbf{2021}, \emph{2}\relax
\mciteBstWouldAddEndPuncttrue
\mciteSetBstMidEndSepPunct{\mcitedefaultmidpunct}
{\mcitedefaultendpunct}{\mcitedefaultseppunct}\relax
\EndOfBibitem
\bibitem[Ali \latin{et~al.}(2022)Ali, Khan, Abbas, and Choi]{ALI2022105844}
Ali,~A.; Khan,~M.~A.; Abbas,~N.; Choi,~H. Prediction of hydrogen storage in dibenzyltoluene empowered with machine learning. \emph{Journal of Energy Storage} \textbf{2022}, \emph{55}, 105844\relax
\mciteBstWouldAddEndPuncttrue
\mciteSetBstMidEndSepPunct{\mcitedefaultmidpunct}
{\mcitedefaultendpunct}{\mcitedefaultseppunct}\relax
\EndOfBibitem
\bibitem[Rahnama \latin{et~al.}(2019)Rahnama, Zepon, and Sridhar]{RAHNAMA20197337}
Rahnama,~A.; Zepon,~G.; Sridhar,~S. Machine learning based prediction of metal hydrides for hydrogen storage, part I: Prediction of hydrogen weight percent. \emph{International Journal of Hydrogen Energy} \textbf{2019}, \emph{44}, 7337--7344\relax
\mciteBstWouldAddEndPuncttrue
\mciteSetBstMidEndSepPunct{\mcitedefaultmidpunct}
{\mcitedefaultendpunct}{\mcitedefaultseppunct}\relax
\EndOfBibitem
\bibitem[Rezakazemi \latin{et~al.}(2017)Rezakazemi, Dashti, Asghari, and Shirazian]{REZAKAZEMI201715211}
Rezakazemi,~M.; Dashti,~A.; Asghari,~M.; Shirazian,~S. H2-selective mixed matrix membranes modeling using ANFIS, PSO-ANFIS, GA-ANFIS. \emph{International Journal of Hydrogen Energy} \textbf{2017}, \emph{42}, 15211--15225\relax
\mciteBstWouldAddEndPuncttrue
\mciteSetBstMidEndSepPunct{\mcitedefaultmidpunct}
{\mcitedefaultendpunct}{\mcitedefaultseppunct}\relax
\EndOfBibitem
\bibitem[Rahnama \latin{et~al.}(2019)Rahnama, Zepon, and Sridhar]{RAHNAMA20197345}
Rahnama,~A.; Zepon,~G.; Sridhar,~S. Machine learning based prediction of metal hydrides for hydrogen storage, part II: Prediction of material class. \emph{International Journal of Hydrogen Energy} \textbf{2019}, \emph{44}, 7345--7353\relax
\mciteBstWouldAddEndPuncttrue
\mciteSetBstMidEndSepPunct{\mcitedefaultmidpunct}
{\mcitedefaultendpunct}{\mcitedefaultseppunct}\relax
\EndOfBibitem
\bibitem[Ahmed \latin{et~al.}(2019)Ahmed, Seth, Purewal, Wong-Foy, Veenstra, Matzger, and Siegel]{ahmed2019exceptional}
Ahmed,~A.; Seth,~S.; Purewal,~J.; Wong-Foy,~A.~G.; Veenstra,~M.; Matzger,~A.~J.; Siegel,~D.~J. Exceptional hydrogen storage achieved by screening nearly half a million metal-organic frameworks. \emph{Nature communications} \textbf{2019}, \emph{10}, 1568\relax
\mciteBstWouldAddEndPuncttrue
\mciteSetBstMidEndSepPunct{\mcitedefaultmidpunct}
{\mcitedefaultendpunct}{\mcitedefaultseppunct}\relax
\EndOfBibitem
\bibitem[Zhou \latin{et~al.}(2022)Zhou, Wang, and Sundmacher]{MOF2}
Zhou,~T.; Wang,~Z.; Sundmacher,~K. \emph{Computer Aided Chemical Engineering}; Elsevier, 2022; Vol.~49; pp 1807--1812\relax
\mciteBstWouldAddEndPuncttrue
\mciteSetBstMidEndSepPunct{\mcitedefaultmidpunct}
{\mcitedefaultendpunct}{\mcitedefaultseppunct}\relax
\EndOfBibitem
\bibitem[Cao \latin{et~al.}(2021)Cao, Dhahad, Zare, Farouk, Anqi, Issakhov, and Raise]{MOF3}
Cao,~Y.; Dhahad,~H.~A.; Zare,~S.~G.; Farouk,~N.; Anqi,~A.~E.; Issakhov,~A.; Raise,~A. Potential application of metal-organic frameworks (MOFs) for hydrogen storage: Simulation by artificial intelligent techniques. \emph{International Journal of Hydrogen Energy} \textbf{2021}, \emph{46}, 36336--36347\relax
\mciteBstWouldAddEndPuncttrue
\mciteSetBstMidEndSepPunct{\mcitedefaultmidpunct}
{\mcitedefaultendpunct}{\mcitedefaultseppunct}\relax
\EndOfBibitem
\bibitem[Sun \latin{et~al.}(2019)Sun, DeJaco, and Siepmann]{sun2019predicting}
Sun,~Y.; DeJaco,~R.~F.; Siepmann,~J.~I. Predicting hydrogen storage in nanoporous materials using meta-learning. Machine Learning and the Physical Sciences Workshop, NeurIPS 2019. 2019\relax
\mciteBstWouldAddEndPuncttrue
\mciteSetBstMidEndSepPunct{\mcitedefaultmidpunct}
{\mcitedefaultendpunct}{\mcitedefaultseppunct}\relax
\EndOfBibitem
\bibitem[Huang \latin{et~al.}(2023)Huang, Magar, Xu, and Farimani]{Huang2023Materials}
Huang,~H.; Magar,~R.; Xu,~C.; Farimani,~A. Materials Informatics Transformer: A Language Model for Interpretable Materials Properties Prediction. \emph{ArXiv} \textbf{2023}, \emph{abs/2308.16259}\relax
\mciteBstWouldAddEndPuncttrue
\mciteSetBstMidEndSepPunct{\mcitedefaultmidpunct}
{\mcitedefaultendpunct}{\mcitedefaultseppunct}\relax
\EndOfBibitem
\bibitem[Fu \latin{et~al.}(2023)Fu, Wei, Song, Li, Xin, Omee, Dong, Siriwardane, and Hu]{fu2023material}
Fu,~N.; Wei,~L.; Song,~Y.; Li,~Q.; Xin,~R.; Omee,~S.~S.; Dong,~R.; Siriwardane,~E. M.~D.; Hu,~J. Material transformers: deep learning language models for generative materials design. \emph{Machine Learning: Science and Technology} \textbf{2023}, \emph{4}, 015001\relax
\mciteBstWouldAddEndPuncttrue
\mciteSetBstMidEndSepPunct{\mcitedefaultmidpunct}
{\mcitedefaultendpunct}{\mcitedefaultseppunct}\relax
\EndOfBibitem
\bibitem[Song \latin{et~al.}(2023)Song, Miret, Zhang, and Liu]{Song2023HoneyBee:}
Song,~Y.; Miret,~S.; Zhang,~H.; Liu,~B. HoneyBee: Progressive Instruction Finetuning of Large Language Models for Materials Science. \textbf{2023}, 5724--5739\relax
\mciteBstWouldAddEndPuncttrue
\mciteSetBstMidEndSepPunct{\mcitedefaultmidpunct}
{\mcitedefaultendpunct}{\mcitedefaultseppunct}\relax
\EndOfBibitem
\bibitem[Zhai \latin{et~al.}(2018)Zhai, Zhang, Chen, and He]{8616075}
Zhai,~J.; Zhang,~S.; Chen,~J.; He,~Q. Autoencoder and Its Various Variants. 2018 IEEE International Conference on Systems, Man, and Cybernetics (SMC). 2018; pp 415--419\relax
\mciteBstWouldAddEndPuncttrue
\mciteSetBstMidEndSepPunct{\mcitedefaultmidpunct}
{\mcitedefaultendpunct}{\mcitedefaultseppunct}\relax
\EndOfBibitem
\bibitem[Mentel()]{mendeleev2014}
Mentel,~L. {mendeleev} -- A Python resource for properties of chemical elements, ions and isotopes. \url{https://github.com/lmmentel/mendeleev}\relax
\mciteBstWouldAddEndPuncttrue
\mciteSetBstMidEndSepPunct{\mcitedefaultmidpunct}
{\mcitedefaultendpunct}{\mcitedefaultseppunct}\relax
\EndOfBibitem
\bibitem[LeCun \latin{et~al.}(2015)LeCun, Bengio, and Hinton]{lecun2015deep}
LeCun,~Y.; Bengio,~Y.; Hinton,~G. Deep learning. \emph{nature} \textbf{2015}, \emph{521}, 436--444\relax
\mciteBstWouldAddEndPuncttrue
\mciteSetBstMidEndSepPunct{\mcitedefaultmidpunct}
{\mcitedefaultendpunct}{\mcitedefaultseppunct}\relax
\EndOfBibitem
\bibitem[Paszke \latin{et~al.}(2019)Paszke, Gross, Massa, Lerer, Bradbury, Chanan, Killeen, Lin, Gimelshein, Antiga, Desmaison, Kopf, Yang, DeVito, Raison, Tejani, Chilamkurthy, Steiner, Fang, Bai, and Chintala]{NEURIPS2019_9015}
Paszke,~A. \latin{et~al.}  \emph{Advances in Neural Information Processing Systems 32}; Curran Associates, Inc., 2019; pp 8024--8035\relax
\mciteBstWouldAddEndPuncttrue
\mciteSetBstMidEndSepPunct{\mcitedefaultmidpunct}
{\mcitedefaultendpunct}{\mcitedefaultseppunct}\relax
\EndOfBibitem
\bibitem[Kingma and Ba(2014)Kingma, and Ba]{kingma2014adam}
Kingma,~D.~P.; Ba,~J. Adam: A method for stochastic optimization. \emph{arXiv preprint arXiv:1412.6980} \textbf{2014}, \relax
\mciteBstWouldAddEndPunctfalse
\mciteSetBstMidEndSepPunct{\mcitedefaultmidpunct}
{}{\mcitedefaultseppunct}\relax
\EndOfBibitem
\bibitem[Ledig \latin{et~al.}(2017)Ledig, Theis, Husz{\'a}r, Caballero, Cunningham, Acosta, Aitken, Tejani, Totz, Wang, \latin{et~al.} others]{ledig2017photo}
Ledig,~C.; Theis,~L.; Husz{\'a}r,~F.; Caballero,~J.; Cunningham,~A.; Acosta,~A.; Aitken,~A.; Tejani,~A.; Totz,~J.; Wang,~Z.; others Photo-realistic single image super-resolution using a generative adversarial network. Proceedings of the IEEE conference on computer vision and pattern recognition. 2017; pp 4681--4690\relax
\mciteBstWouldAddEndPuncttrue
\mciteSetBstMidEndSepPunct{\mcitedefaultmidpunct}
{\mcitedefaultendpunct}{\mcitedefaultseppunct}\relax
\EndOfBibitem
\bibitem[Kresse and Furthmüller(1996)Kresse, and Furthmüller]{Kresse1996CMS}
Kresse,~G.; Furthmüller,~J. Efficiency of ab-initio total energy calculations for metals and semiconductors using a plane-wave basis set. \emph{Computational Materials Science} \textbf{1996}, \emph{6}, 15--50\relax
\mciteBstWouldAddEndPuncttrue
\mciteSetBstMidEndSepPunct{\mcitedefaultmidpunct}
{\mcitedefaultendpunct}{\mcitedefaultseppunct}\relax
\EndOfBibitem
\bibitem[Kresse and Furthm\"uller(1996)Kresse, and Furthm\"uller]{Kresse1996PRB}
Kresse,~G.; Furthm\"uller,~J. Efficient iterative schemes for ab initio total-energy calculations using a plane-wave basis set. \emph{Phys. Rev. B} \textbf{1996}, \emph{54}, 11169--11186\relax
\mciteBstWouldAddEndPuncttrue
\mciteSetBstMidEndSepPunct{\mcitedefaultmidpunct}
{\mcitedefaultendpunct}{\mcitedefaultseppunct}\relax
\EndOfBibitem
\bibitem[Kresse and Joubert(1999)Kresse, and Joubert]{Kresse1999}
Kresse,~G.; Joubert,~D. From ultrasoft pseudopotentials to the projector augmented-wave method. \emph{Phys. Rev. B} \textbf{1999}, \emph{59}, 1758--1775\relax
\mciteBstWouldAddEndPuncttrue
\mciteSetBstMidEndSepPunct{\mcitedefaultmidpunct}
{\mcitedefaultendpunct}{\mcitedefaultseppunct}\relax
\EndOfBibitem
\bibitem[Bl\"ochl(1994)]{Bl1994}
Bl\"ochl,~P.~E. Projector augmented-wave method. \emph{Phys. Rev. B} \textbf{1994}, \emph{50}, 17953--17979\relax
\mciteBstWouldAddEndPuncttrue
\mciteSetBstMidEndSepPunct{\mcitedefaultmidpunct}
{\mcitedefaultendpunct}{\mcitedefaultseppunct}\relax
\EndOfBibitem
\bibitem[Grimme(2006)]{Grimme2006}
Grimme,~S. Semiempirical GGA-type density functional constructed with a long-range dispersion correction. \emph{Journal of Computational Chemistry} \textbf{2006}, \emph{27}, 1787--1799\relax
\mciteBstWouldAddEndPuncttrue
\mciteSetBstMidEndSepPunct{\mcitedefaultmidpunct}
{\mcitedefaultendpunct}{\mcitedefaultseppunct}\relax
\EndOfBibitem
\bibitem[Perdew \latin{et~al.}(1996)Perdew, Burke, and Ernzerhof]{Perdew1996}
Perdew,~J.~P.; Burke,~K.; Ernzerhof,~M. Generalized Gradient Approximation Made Simple. \emph{Phys. Rev. Lett.} \textbf{1996}, \emph{77}, 3865--3868\relax
\mciteBstWouldAddEndPuncttrue
\mciteSetBstMidEndSepPunct{\mcitedefaultmidpunct}
{\mcitedefaultendpunct}{\mcitedefaultseppunct}\relax
\EndOfBibitem
\bibitem[Sullivan(2000)]{doi:https://doi.org/10.1002/0471238961.0825041819211212.a01}
Sullivan,~E.~A. \emph{Kirk‐Othmer Encyclopedia of Chemical Technology}; John Wiley \& Sons, Ltd, 2000\relax
\mciteBstWouldAddEndPuncttrue
\mciteSetBstMidEndSepPunct{\mcitedefaultmidpunct}
{\mcitedefaultendpunct}{\mcitedefaultseppunct}\relax
\EndOfBibitem
\bibitem[Tibshirani(1996)]{tibshirani1996regression}
Tibshirani,~R. Regression shrinkage and selection via the lasso. \emph{Journal of the Royal Statistical Society Series B: Statistical Methodology} \textbf{1996}, \emph{58}, 267--288\relax
\mciteBstWouldAddEndPuncttrue
\mciteSetBstMidEndSepPunct{\mcitedefaultmidpunct}
{\mcitedefaultendpunct}{\mcitedefaultseppunct}\relax
\EndOfBibitem
\bibitem[Friedman(2001)]{friedman2001greedy}
Friedman,~J.~H. Greedy function approximation: a gradient boosting machine. \emph{Annals of statistics} \textbf{2001}, 1189--1232\relax
\mciteBstWouldAddEndPuncttrue
\mciteSetBstMidEndSepPunct{\mcitedefaultmidpunct}
{\mcitedefaultendpunct}{\mcitedefaultseppunct}\relax
\EndOfBibitem
\bibitem[Chen \latin{et~al.}(2021)Chen, Li, and Wu]{Chen_2021}
Chen,~P.; Li,~F.; Wu,~C. Research on Intrusion Detection Method Based on Pearson Correlation Coefficient Feature Selection Algorithm. \emph{Journal of Physics: Conference Series} \textbf{2021}, \emph{1757}, 012054\relax
\mciteBstWouldAddEndPuncttrue
\mciteSetBstMidEndSepPunct{\mcitedefaultmidpunct}
{\mcitedefaultendpunct}{\mcitedefaultseppunct}\relax
\EndOfBibitem
\bibitem[Das \latin{et~al.}(2023)Das, Thekkepat, Lee, Lee, and Bhattacharjee]{das2023computational2}
Das,~P.; Thekkepat,~K.; Lee,~Y.-S.; Lee,~S.-C.; Bhattacharjee,~S. Computational design of novel MAX phase alloys as potential hydrogen storage media combining first principles and cluster expansion methods. \emph{Physical Chemistry Chemical Physics} \textbf{2023}, \emph{25}, 5203--5210\relax
\mciteBstWouldAddEndPuncttrue
\mciteSetBstMidEndSepPunct{\mcitedefaultmidpunct}
{\mcitedefaultendpunct}{\mcitedefaultseppunct}\relax
\EndOfBibitem
\bibitem[Wolf \latin{et~al.}(2020)Wolf, Debut, Sanh, Chaumond, Delangue, Moi, Cistac, Rault, Louf, Funtowicz, Davison, Shleifer, von Platen, Ma, Jernite, Plu, Xu, Scao, Gugger, Drame, Lhoest, and Rush]{huggingface}
Wolf,~T. \latin{et~al.}  HuggingFace's Transformers: State-of-the-art Natural Language Processing. \url{https://huggingface.co}, 2020; Accessed: 2024-06-03\relax
\mciteBstWouldAddEndPuncttrue
\mciteSetBstMidEndSepPunct{\mcitedefaultmidpunct}
{\mcitedefaultendpunct}{\mcitedefaultseppunct}\relax
\EndOfBibitem
\bibitem[Vaswani \latin{et~al.}(2017)Vaswani, Shazeer, Parmar, Uszkoreit, Jones, Gomez, Kaiser, and Polosukhin]{vaswani2017attention}
Vaswani,~A.; Shazeer,~N.; Parmar,~N.; Uszkoreit,~J.; Jones,~L.; Gomez,~A.~N.; Kaiser,~{\L}.; Polosukhin,~I. Attention is all you need. \emph{Advances in neural information processing systems} \textbf{2017}, \emph{30}\relax
\mciteBstWouldAddEndPuncttrue
\mciteSetBstMidEndSepPunct{\mcitedefaultmidpunct}
{\mcitedefaultendpunct}{\mcitedefaultseppunct}\relax
\EndOfBibitem
\bibitem[Kusaba \latin{et~al.}(2022)Kusaba, Liu, and Yoshida]{KUSABA2022111496}
Kusaba,~M.; Liu,~C.; Yoshida,~R. Crystal structure prediction with machine learning-based element substitution. \emph{Computational Materials Science} \textbf{2022}, \emph{211}, 111496\relax
\mciteBstWouldAddEndPuncttrue
\mciteSetBstMidEndSepPunct{\mcitedefaultmidpunct}
{\mcitedefaultendpunct}{\mcitedefaultseppunct}\relax
\EndOfBibitem
\end{mcitethebibliography}

\end{document}